\documentclass[journal]{IEEEtran}

\ifCLASSINFOpdf
\else
\fi

\usepackage{comment}
\usepackage[nolist,nohyperlinks]{acronym}
\usepackage{url}
\usepackage{booktabs}
\usepackage{amsmath}
\usepackage{textcomp}
\usepackage{graphicx}
\graphicspath{ {./images/} }
\usepackage{etoolbox}
\makeatletter
\pretocmd{\@todo}{\GenericWarning{}{There's sometthing to do here}}{}{}
\makeatother
\usepackage{cite}
\usepackage{amssymb}
\usepackage{amsfonts}
\usepackage[caption=false]{subfig}
\usepackage{booktabs}
\usepackage[colorinlistoftodos,prependcaption]{todonotes}
\usepackage{xcolor}
\usepackage{pifont}
\usepackage{tabularx,colortbl}
\usepackage{cite}
\usepackage{amssymb}
\usepackage[]{hyperref}
\usepackage{array}
\newcolumntype{H}{>{\setbox0=\hbox\bgroup}c<{\egroup}@{}}
\newcommand{\cmark}{\ding{51}}%
\newcommand{\xmark}{\ding{55}}%
\usepackage{stfloats} %
\usepackage{multirow}

\acrodef{TFAN}{Time-Frequency Adaptive Normalization}
\acrodef{ASV}{Automatic Speaker Verification}
\acrodef{TF}{Time-Frequency}
\acrodef{GAN}{Generative Adversarial Network}
\acrodef{CGAN}{Conditional GAN}
\acrodef{CycleGAN}{Cycle-consistent GAN}
\acrodef{BWE}{Bandwidth Extension}
\acrodef{STFT}{Short-Time Fourier Transform}
\acrodef{MRSTFT}{Multi-resolution Short-Time Fourier Transform}
\acrodef{cGAN}{Conditional Generative Adversarial Network}
\acrodef{MAE}{Mean Absolute Error}
\acrodef{MSE}{Mean Squared Error}
\acrodef{TTUR}{Two-Time scale Update Rule}
\acrodef{UNB}{Upsampled Narrowband}
\acrodef{WB}{Wideband}
\acrodef{NB}{Narrowband}
\acrodef{STMN}{Short Time Mean Normalization}
\acrodef{SSDA}{Single-source Domain Adaptation}
\acrodef{MSDA}{Multi-source Domain Adaptation}
\acrodef{SNR}{Signal-to-Noise Ratio}
\acrodef{SINR}{Signal-to-Interference-plus-Noise Ratio}
\acrodef{DNN}{Deep Neural Network}
\acrodef{TTS}{Text-to-Speech}
\acrodef{TFNet}{Time-frequency Network}
\acrodef{MFCC}{Mel-Filter Cepstral Coefficient}
\acrodef{DFL}{Deep Feature Loss}
\acrodef{DDPM}{Denoising Diffusion Probabilistic Model}
\acrodef{CNN}{Convolutional Neural Network}
\acrodef{FiLM}{Feature-wise Linear Modulation}
\acrodef{AFiLM}{Attention-based Feature-wise Linear Modulation}
\acrodef{TFiLM}{Temporal Feature-wise Linear Modulation}
\acrodef{LPC}{Linear Predictive Coding}
\acrodef{GMM}{Gaussian Mixture Model}
\acrodef{HMM}{Hidden Markov Model}
\acrodef{PLDA}{Probabilistic Linear Discriminant Analysis}
\acrodef{DAT}{Domain Adversarial Training}
\acrodef{DAE}{Denoising Auto-Encoder}
\acrodef{HFE}{High Frequency Energy}
\acrodef{BLSTM}{Bi-directional Long-Short Term Memory}
\acrodef{ASR}{Automatic Speech Recognition}
\acrodef{FM}{Feature Matching}
\acrodef{AFM}{Auxiliary Feature Matching}
\acrodef{WER}{Word Error Rate}
\acrodef{EER}{Equal Error Rate}
\acrodef{minDCF}{Minimum Decision Cost Function}
\acrodef{SRE}{Speaker Recognition Evaluation}
\acrodef{SOTA}{state-of-the-art}
\acrodef{LSGAN}{Least Squares GAN}
\acrodef{AGD}{Alternating Gradient Descent}
\acrodef{UB}{upperband}
\acrodef{LB}{lowerband}
\acrodef{UPR}{upsampling ratio}
\acrodef{LSD}{Log-Spectral Distance}
\acrodef{PESQ}{Perceptual Evaluation of Speech Quality}
\acrodef{STOI}{Short-Time Objective Intelligibility}
\acrodef{VAE}{Variational Auto-Encoder}

\begin{document}
\title{Time-domain speech super-resolution with GAN based modeling for telephony speaker verification}

\author{Saurabh Kataria,~\IEEEmembership{Student Member,~IEEE}, Jes\'us Villalba,~\IEEEmembership{Member,~IEEE}, Laureano Moro-Vel\'azquez,~\IEEEmembership{Member,~IEEE}, Piotr \.Zelasko,~\IEEEmembership{Member,~IEEE}, Najim Dehak,~\IEEEmembership{Senior Member,~IEEE}
\thanks{All authors are associated with the Department of Electrical and Computer Engineering and Center for Language and Speech Processing (CLSP), Johns Hopkins University, Baltimore, MD, USA, 21218.

Saurabh Kataria, Jes\'us Villalba, Najim Dehak are also associated with Human Language Technology Center of Excellence (HLTCOE), Johns Hopkins University, Baltimore, MD, USA, 21218. Piotr \.Zelasko is current at Meaning.}%
\thanks{}%
\thanks{}}

\markboth{}%
{Shell \MakeLowercase{\textit{et al.}}: Bare Demo of IEEEtran.cls for IEEE Journals}
\markboth{Journal of \LaTeX\ Class Files,~Vol.~XX, No.~XX, September~2022}%
{Shell \MakeLowercase{\textit{et al.}}: Bare Demo of IEEEtran.cls for IEEE Journals}

\maketitle

\begin{abstract}
Automatic Speaker Verification (ASV) technology has become commonplace in virtual assistants.
However, its performance suffers when there is a mismatch between the train and test domains.
Mixed bandwidth training, i.e., pooling training data from both domains, is a preferred choice for developing a universal model that works for both narrowband and wideband domains.
We propose complementing this technique by performing neural upsampling of narrowband signals, also known as bandwidth extension.
Our main goal is to discover and analyze high-performing time-domain Generative Adversarial Network (GAN) based models to improve our downstream state-of-the-art ASV system.
We choose GANs since they (1) are powerful for learning conditional distribution and (2) allow flexible \emph{plug-in} usage as a pre-processor during the training of downstream task (ASV) with data augmentation.
Prior works mainly focus on feature-domain bandwidth extension and limited experimental setups.
We address these limitations by 1) using time-domain extension models, 2) reporting results on three real test sets, 2) extending training data, and 3) devising new test-time schemes.
We compare supervised (conditional GAN) and unsupervised GANs (CycleGAN) and demonstrate average relative improvement in Equal Error Rate of 8.6\% and 7.7\%, respectively.
For further analysis, we study changes in spectrogram visual quality, audio perceptual quality, t-SNE embeddings, and ASV score distributions.
We show that our bandwidth extension leads to phenomena such as a shift of telephone (test) embeddings towards wideband (train) signals, a negative correlation of perceptual quality with downstream performance, and condition-independent score calibration.
\end{abstract}

\begin{IEEEkeywords}
bandwidth extension, speaker verification, conditional GAN, CycleGAN, perceptual quality, score distribution
\end{IEEEkeywords}

\IEEEpeerreviewmaketitle

\vspace{-2mm}
\section{Introduction}
\label{sec:intro}
\IEEEPARstart{S}{peech} technologies such as voice assistants have proliferated recently, thanks to the advancements made in deep learning~\cite{yu2016automatic,bai2021speaker,tan2021survey}.
Usually they are designed for a particular acoustic environment, which causes a mismatch between the train and test data in terms of channel, acoustic domain, sampling frequency, and \ac{SNR}~\cite{garcia2019speaker}.
There is also a degradation in performance of downstream tasks like \ac{ASR}~\cite{maas2012recurrent} and \ac{ASV}~\cite{villalba2019state,kataria2020feature,kataria2020analysis}.
It is challenging to develop a universal model invariant to the choice of testing \emph{domains} like narrowband telephone speech, far-field speech, and children's speech in the wild~\cite{garcia2019speaker}.
Common techniques to promote domain invariance include \ac{DAT}~\cite{wang2021multi}, mixed-bandwidth training~\cite{cai2020unified}, multi-task learning~\cite{hou2020multi,kataria2022joint}, feeding auxiliary information~\cite{mantena2019bandwidth}, and pre-processing solutions like speech enhancement~\cite{kataria2020feature}, bandwidth extension~\cite{nidadavolu2018investigation,kataria2022joint}.
In this paper, we focus on \ac{BWE} for the downstream task of telephony ASV, where the goal is to determine whether speakers in two given recordings are identical or not -- primarily in telephone test sets.
Typically, \ac{ASV} is trained on mixed bandwidth data i.e. \ac{NB} telephone and \ac{WB} microphone which are bandlimited to 4~KHz and 8~KHz, respectively.
We focus on bandwidth mismatch through neural upsampling, also known as bandwidth extension/expansion, audio super-resolution, or simply \emph{extension}.
It refers to increasing the bandwidth, i.e., the highest frequency information available) in \ac{NB} signals to match with the bandwidth of \ac{WB} signals.
Equivalently, it amounts to estimating the missing \ac{UB} frequency region (also known as \ac{HFE}) from the \ac{LB} region.
The factor by which the bandwidth increases is called \ac{UPR}~\cite{liu2022neural}.

Active research in BWE has improved the quality of extended signals over the years.
Modeling domain choices include time and frequency, where the latter is more prevalent in the past.
A combination of the two approaches is also prevalent.
In the time-domain approach, BWE is performed directly on temporal samples.
In the frequency-domain approach, we predict only the \ac{STFT} magnitude of the speech signal, while we re-use the old phase (only for the lower band). %

BWE literature is also categorized based on the usage of generative modeling.
For the evaluation of extended signals, human listening studies~\cite{li2015dnn} and distortion metrics such as \ac{LSD}~\cite{gu2016speech} are used.
Arguably, perceptual and intelligibility metrics of speech enhancement metrics like \ac{PESQ}~\cite{rix2001perceptual}, \ac{STOI}~\cite{taal2011algorithm} can be used as well.
BWE can also improve speech-in-noise perception, source localization, speech intelligibility, gender identification, and phoneme identification~\cite{monson2014perceptual,donai2017gender,vitela2015phoneme}.
Most prior works study BWE independently, but we can also pursue it with other tasks: 1) joint learning of BWE with other tasks to obtain a better BWE model or 2) learning BWE to use it as a pre-processor for improving downstream tasks.
The second direction concerns downstream performance explicitly.
We referred to this approach as \emph{task-specific} enhancement~\cite{kataria2020feature,kataria2020analysis}.

State-of-the-art speaker verification systems use x-vector~\cite{snyder2018x} as front-end for embedding extraction and \ac{PLDA}~\cite{kenny2010bayesian} as back-end for scoring and evaluation~\cite{villalba2019state}.
We choose a robust experimental setup (Sec.~\ref{sec:exp}) where we train ASV with narrowband as well as wideband data as per mixed bandwidth training protocol~\cite{nidadavolu2018investigation}.
We use data augmentation by adding noise and reverberation during training.
To train ASV with bandwidth expanded samples and data augmentation, we design time-domain BWE systems since augmentation allows flexible plug-in usage.
We wish to analyze the effect of BWE via 1) ASV scores, 2) speaker embedding, and 3) perceptual quality.
Typically, we measure the benefit of extension on downstream tasks via improvement averaged across test condition types.
When test condition information is available, fine-grained analysis is also possible.
Furthermore, changes in \emph{target/non-target} ASV trial score distribution can reveal interesting properties of extension.
We can also study the shift of speaker embeddings of the test set w.r.t. train set to reveal the domain adaptation capability of extension.
The extension can also modify the perceptual quality of speech, and we study its correlation with downstream performance using supervised and unsupervised GANs.
We provide the first comprehensive work that:
\begin{enumerate}
    \item Discusses in-depth how to design a strong time-domain GAN-based bandwidth extension system for \ac{ASV}.
    \item Provides fair comparison of supervised (deep regression, CGAN) and unsupervised (CycleGAN) methods on three real test sets.
    \item Comprehensively evaluates of BWE system by extending training data of ASV (PLDA and x-vector).
    \item Analyzes in detail analysis via per-trial type reporting, spectrogram visualization, automated perceptual quality assessment, and embedding visualization for extended signals -- for both supervised and unsupervised methods.
\end{enumerate}

\vspace{-2mm}
\section{Prior Work}
Traditional approaches to bandwidth extension included source-filter model~\cite{makhoul1979high}, \ac{GMM}~\cite{seo2014maximum}, \ac{LPC}~\cite{bachhav2018efficient}, and \ac{HMM}~\cite{jax2003artificial}.
Advances in deep learning improved the modeling power significantly.
Deep regression~\cite{li2015deep} (\emph{mapping}) became one of the earliest successful techniques for BWE and speech enhancement.
Prior works have explored various modeling domains, architecture, objective function, and paradigm.
Time-domain models offer maximum flexibility. However, they have only recently become as performing as their frequency-domain counterpart.
\cite{kuleshov2017audio} develops BWE in time-domain using a simple $l_2$ (\ac{MSE}) regression loss.
\cite{li2019speech} is a critical relevant work where authors used \ac{DFL} in addition to time and frequency domain losses.
Some studies utilize re-use of the initial phase in the frequency-domain network such that output is temporal~\cite{lin2021two,hu2020phase}.
Typically, the choice of modeling domain and architecture are interdependent.
For time-domain systems, mechanisms like 1-D CNNs are popular.
In \cite{nguyen2021tunet}, authors trained time-domain system using \ac{TFiLM} mechanism and \emph{deep feature loss}~\cite{feng2019learning,kataria2020feature,kataria2020analysis,kataria2021perceptual,kataria2021deep}, whose performance is further boosted by self-supervised pre-training and data augmentation.
Although we do not pursue pre-training, we employ data augmentation in PLDA and x-vector networks to use more robust and realistic baselines.

GANs are a popular choice of generative model for BWE.
Supervised \ac{GAN} in form of \ac{CGAN} was used in \cite{su2021bandwidth}.
The main limitations were the lack of downstream task evaluation and the limited exploration of GAN parameters, which we address in our work.
A proof-of-concept for using unsupervised GAN called CycleGAN (cycle-consistent GAN) exists in \cite{haws2019cyclegan}.
Authors report improvement for the downstream task of \ac{ASR} by learning upper-band spectral coefficients, although the improvement is minimal, and the modeling is not in the time domain.
We address these limitations too.
Using the discriminator model for adversarial learning is called \emph{feature-matching} loss.
We also experiment using an external model and term the loss Auxiliary Feature Matching (AFM) loss.

BWE can be pursued together with other tasks as well.
\cite{hou2020multi} employs a multi-task framework of BWE with denoising.
Another relevant work for extending historical recordings is \cite{moliner2022behm}.
We do not pursue multi-task learning.
However, in \cite{kataria2022joint}, we extend our work to joint learning with domain adaptation.
BWE can help improve downstream tasks like speaker recognition and speech recognition.
In \cite{yamamoto2019speaker}, authors used BWE as a pre-processing step for narrowband data for mixed-bandwidth training of a robust wideband speaker embedding network.
Another type of mixed-bandwidth training~\cite{cai2020unified} is where narrowband and wideband speaker identities are classified separately while using the same feature extractor backbone.
Previous works have shown that BWE of training data improves ASV~\cite{nidadavolu2018investigation,snyder2018x,nidadavolu2019investigation}.
In ASR, BWE is shown to improve performance in terms of Word Error Rate~\cite{li2015dnn}.

Log-Spectral Distortion (LSD)~\cite{nidadavolu2019investigation} is commonly used to measure distortion in BWE outputs in the frequency domain.
We wish to use other metrics like simple time-domain Mean Squared Error (MSE) and, more importantly, speech enhancement metrics.
For the output spectrogram quality, prior works~\cite{sivaraman2020speech,li2015deep,gupta2019speech} revealed 1) predicted upper band does not have enough energy and appears extended trivially from lower band voiced frames, 2) discontinuity at the intersection of the lower band and upper band.
In our work, we investigate if GANs can rectify these.
In \cite{kataria2021deep}, authors showed that, through t-distributed Stochastic Neighbour Embedding (t-SNE) analysis, (speaker) embeddings of \emph{source} speech can shift towards \emph{target} in domain adaptation application.
We are interested in such an analysis for BWE.
To our knowledge, ASV score analysis is not done in the past. %

\section{Bandwidth Extension Models}
\label{sec:bwe_models}
This section describes the three models we use for bandwidth extension: deep regression, conditional GAN, and CycleGAN, along with their objective function and architectures.

\subsection{Deep regression}
Also known as \emph{mapping approach}~\cite{li2015deep}, this involves training a feedforward network to directly predict the desired output via regression loss. 
This technique requires paired data.
The input would be narrowband data, and the target would be the corresponding (paired) wideband audio.

\subsection{Conditional GAN}
\label{sec:cgan}
GANs~\cite{goodfellow2014generative} are effective in sampling from the true data distribution.
Conditional GANs~\cite{mirza2014conditional} are supervised variant of GAN which learns to sample from conditional distribution and, thus, can map samples from domain A to B.
The generator $\mathcal{G}_{A\rightarrow B}$ learns this mapping while discriminator $\mathcal{D}_{B}$ learns to distinguish between fake (generated) and real via
\begin{align}
\label{eq:cgan}
    \max_{\mathcal{G}_{A\rightarrow B}} &\min_{\mathcal{D}_{B}} \mathcal{L}_{\text{CGAN}} \hspace{0.5em} \text{, where} \\
        \mathcal{L}_{\text{CGAN}} &= \mathcal{L}_{\text{adv}} + \lambda_{\text{sup}}\mathcal{L}_{\text{sup}}.
\end{align}
Here, $\mathcal{L}_{\text{adv}}$ is the adversarial loss, and ($\mathbf{a}$, $\mathbf{b}$ are (paired) samples from the distribution of A and B domains, respectively.
The supervised loss $\mathcal{L}_{\text{sup}}$ using $\mathcal{L}_p$ norm is given by
\begin{equation}
\label{eq:cgan_sup}
    \mathcal{L}_{\text{sup}} = \mathbb{E}_{\mathbf{a}\sim p_A, \mathbf{b}\sim p_B} [\lVert\mathbf{b} - \mathcal{G}_{A\rightarrow B}(\mathbf{a})\rVert_p].
\end{equation}

\subsection{CycleGAN}
\label{sec:cyclegan}
Cycle-consistent GAN is an unsupervised GAN that learns mappings between two domains~\cite{zhu2017unpaired}.
In addition to two adversarial losses, it has cycle loss $\mathcal{L}_{\text{cyc}}$ and identity loss $\mathcal{L}_{\text{id}}$, weighted by hyper-parameters $\lambda_{\text{cyc}}$ and $\lambda_{\text{id}}$ respectively.
The role of cycle loss is to enforce semantic consistency during translation/mapping, while identity loss serves as a regularizer.
Let $\mathcal{G}_{A\rightarrow B}$, $\mathcal{G}_{B\rightarrow A}$ denote the generators and $\mathcal{D}_{A}$, $\mathcal{D}_{B}$ denote the discriminators.
We optimize
\begin{align}
\label{eq:cyclegan}
    &\max_{\mathcal{G}_{A\rightarrow B}, \mathcal{G}_{B\rightarrow A}} \min_{\mathcal{D}_{A}, \mathcal{D}_{B}} \mathcal{L}_{\text{cyc-GAN}} \hspace{0.5em} \text{, where} \hspace{2em}\\
        \mathcal{L}_{\text{cyc-GAN}} &= \mathcal{L}_{\text{adv},A\rightarrow B} + \mathcal{L}_{\text{adv},B\rightarrow A} +
        \lambda_{\text{cyc}}\mathcal{L}_{\text{cyc}} + \lambda_{\text{id}}\mathcal{L}_{\text{id}}.
\end{align}
Adversarial losses $\mathcal{L}_{\text{adv},A\rightarrow B}$ and $\mathcal{L}_{\text{adv},B\rightarrow A}$ are defined in Sec.~\ref{sec:adv_losses}.
Cycle and identity losses are
\begin{align}
    \mathcal{L}_{\text{cyc}} &= \mathbb{E}_{\mathbf{a}\sim p_{A},\mathbf{b}\sim p_{B}} [\lVert\mathbf{a} - \mathcal{G}_{B\rightarrow A}(\mathcal{G}_{A\rightarrow B}(\mathbf{a}))\rVert_1]\notag\\
    &+ \mathbb{E}_{\mathbf{a}\sim p_{A},\mathbf{b}\sim p_{B}} [\lVert\mathbf{b} - \mathcal{G}_{A\rightarrow B}(\mathcal{G}_{B\rightarrow A}(\mathbf{b}))\rVert_1],\\
    \mathcal{L}_{\text{id}} &= \mathbb{E}_{\mathbf{a}\sim p_{A}}[\lVert\mathbf{a} - \mathcal{G}_{B\rightarrow A}(\mathbf{a})\rVert_1] \notag \\
    &+ \mathbb{E}_{\mathbf{b}\sim p_{B}}[\lVert\mathbf{b} - \mathcal{G}_{A\rightarrow B}(\mathbf{b})\rVert_1].
\end{align}
Since there is no direct supervision loss in CycleGAN, it does not require paired samples from two domains for training and, hence, offers more flexibility in usage compared with CGAN, although at the cost of increased complexity.
$(\mathbf{a},\mathbf{b})$ represents unpaired data.
There does not necessarily exist any relation between $\mathbf{a}$ and $\mathbf{b}$.
However, we can use paired data instead and term it \emph{paired CycleGAN}.

\begin{figure*}[htbp]
\centering
\includegraphics[trim={0.5cm 0.5cm 0.5cm 0},clip,width=0.92\textwidth]{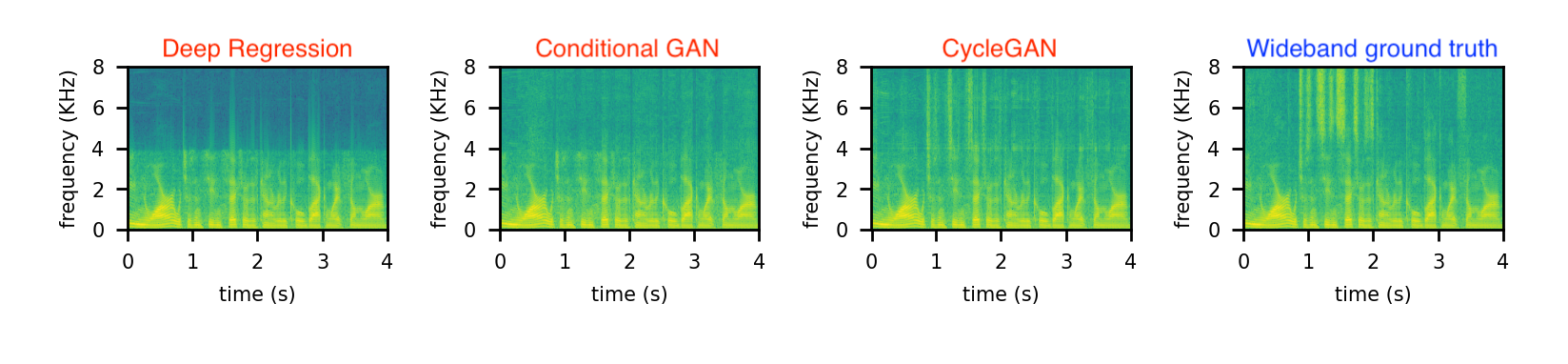}
\vspace{-0.45em}
\includegraphics[width=0.92\textwidth]{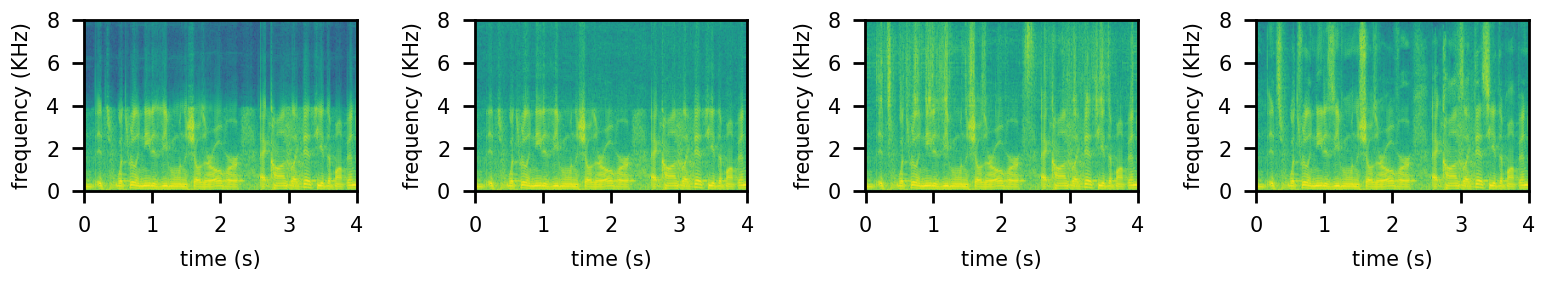}
\vspace{-0.45em}
\includegraphics[width=0.92\textwidth]{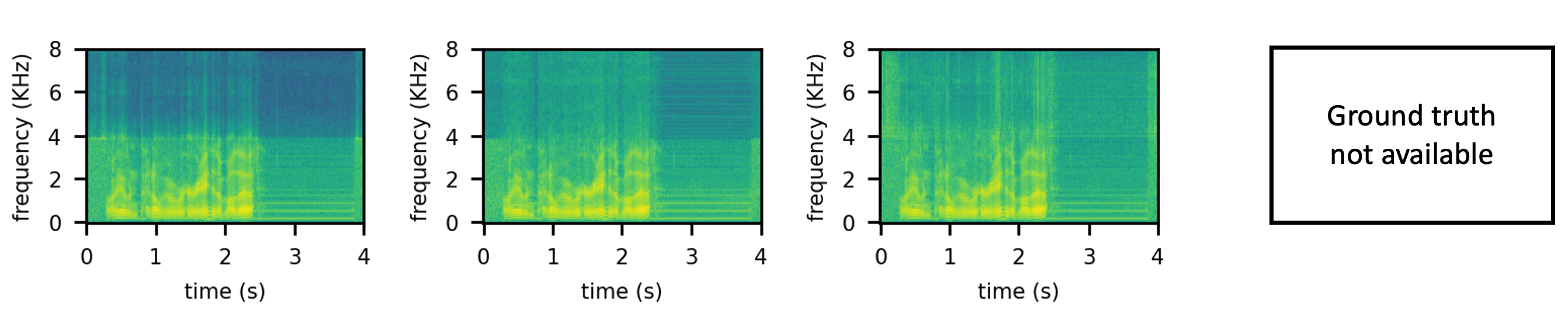}
\vspace{-0.45em}
\includegraphics[width=0.92\textwidth]{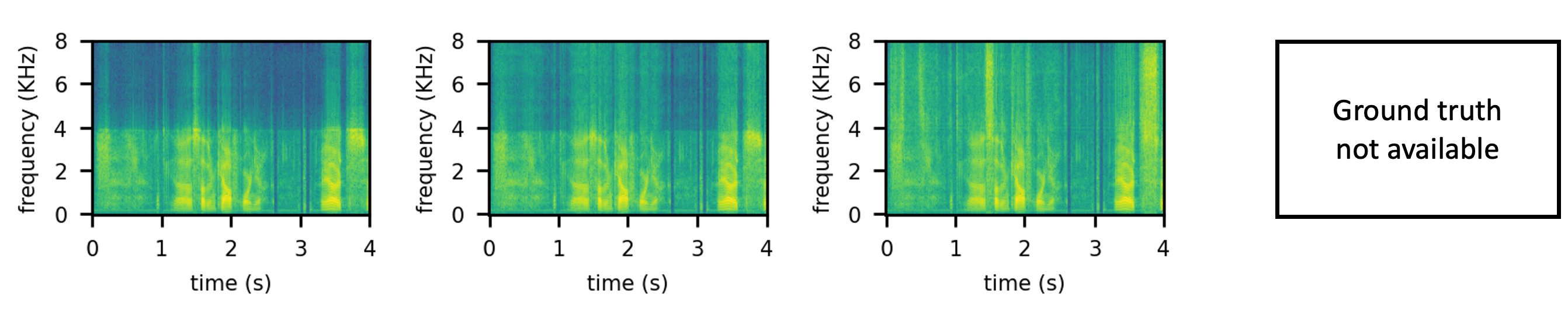}
\caption{Spectrograms for four random samples (one per row) from SRE21 test set after extension by deep regression, CGAN, and CycleGAN in columns 1, 2, and 3, respectively.
The last column contains the ground truth wideband spectrogram.
It is unavailable for the last two samples since they are originally narrowband signals unlike the other two samples.
Note that GAN (especially CycleGAN) produces the best (natural-looking) outcomes.
}
\label{fig:spectrograms}
\vspace{-5mm}
\end{figure*}

\subsection{Generator architecture}
\label{sec:gen_arch}
\textbf{Conv-TasNet}~\cite{luo2019conv}:
We use off-the-shelf Conv-TasNet for the feedforward model for deep regression and the generator for GANs.
It is a time-domain model that has been used for other tasks like speech enhancement~\cite{joshi2022advest,joshi2022defense} and source separation~\cite{luo2019conv}.
It consists of \emph{encoder}, \emph{separator}, and \emph{decoder}.
The \emph{encoder} is simply a 1-D Convolutional Neural Network (CNN) layer.
The \emph{separator}, through multiple 1-D CNN layers, computes a mask for separation used by \emph{decoder} stage to produce output.
We make minor modifications in the architecture~\footnote{\url{https://github.com/naplab/Conv-TasNet}}.
The number of stacks of CNNs in \emph{separator} is one, the number of layers per stack is eight, and the number of output channels of \emph{decoder} is one.
To capture finer details in high sampling rate signals, we choose the number of channels in \emph{encoder} as 128, the kernel size as 16, and the stride as 8.
The number of input and output channels in the separator are 128 and 1024, respectively, and dilation increases exponentially with a factor of 2.
The receptive field is 32~ms and the total parameters are only 1.6~M making the deployment lightweight.

\subsection{Discriminator architectures}
\label{sec:disc_archs}

\textbf{Parallel WaveGAN (PWG)}~\cite{yamamoto2020parallel}:
This is a 1-D deep CNN with ten CNN layers with a kernel size of 3, channels as 80, activation as LeakyReLU (slope=-0.2), and linearly increasing dilation from the second layer to the ninth layer (from dilation value of one to eight).
It contains only 0.16M parameters.

\textbf{MelGAN}~\cite{kumar2019melgan}:
This model is similar to PWG, and works well with a variety of adversarial losses.
There are seven CNN layers and the corresponding channels are \{16,64,256,1024,1024,1024,1\} and kernel sizes are \{15,41,41,41,41,5,3\}.
The number of parameters is 5.6M, thus significantly bigger than PWG.
We also use a \emph{multi-scale} version of MelGAN, which uses three such discriminators.
Here, \emph{scale} refers to the downsampling factor used for the input to the model.
We use the same scale value of four for all sub-discriminators.

\textbf{HiFiGAN}~\cite{kong2020hifi}:
We use its multi-scale and multi-period versions.
\emph{Period} parameter signifies splitting signal (with length equal to period) and concatenating the parts along a new axis.
The multi-period model consists of four discriminators.
Each consists of six 1-D CNN layers with kernel size as 5, stride as 3, output channels as \{4, 16, 64, 256, 1024, 1\}, and LeakyReLU activation (slope=-0.1).
The number of parameters is 7M.
The multi-scale model consists of three discriminators each containing eight CNN layers with kernel size as \{15, 41, 41, 41, 41, 41, 5, 3\}, stride as \{1, 2, 2, 4, 4, 1, 1, 1\}, output channels as \{16, 16, 32, 64, 128, 256, 512, 1\}, and LeakyReLU activation (slope=-0.1).
The multi-scale multi-period model uses both models simultaneously with total 2.3M parameters.

\textbf{StyleMelGAN}~\cite{mustafa2021stylemelgan}:
This consists of differentiable Pseudo Quadrature Mirror Filter bank (PQMF) analysis as pre-processing for input to four sub-models, each analyzing a different signal subband.
Each sub-model is a MelGAN discriminator.
The number of parameters is 5.9M.

\vspace{-2mm}
\subsection{Supervision losses}
\label{sec:sup_losses}
For the supervision loss in CGAN and identity and cycle loss in CycleGAN, we use the following loss functions: simple MSE ($l_2$) and MAE ($l_1$) losses.

\textbf{Multi-Resolution Short-Time Fourier Transform} \cite{yamamoto2020parallel}:
MRSTFT loss simply compares two signals $\mathbf{x}$ and $\mathbf{x'}$ in frequency domain by using $M=3$ STFT with corresponding FFT sizes (N) as \{1024, 2048, 512\}, hop sizes as \{120, 240, 50\}, window length as \{600, 1200, 240\}, and Hann window:
\begin{align}
\label{eq:mrstft}
\mathcal{L}_{\text{sup}} &= \mathbb{E}_{\mathbf{x,x'}}[\sum_{m=1}^M \mathcal{L}_{\text{sup}}^{(m)}(\mathbf{x},\mathbf{x'})]\;,\\
    \mathcal{L}_{\text{sup}}^{(m)}(\mathbf{x},\mathbf{x'}) &=  \mathcal{L}_{\text{sc}}^{(m)}(\mathbf{x},\mathbf{x'}) 
    + \mathcal{L}_{\text{mag}}^{(m)}(\mathbf{x},\mathbf{x'}),\\
    \mathcal{L}_{\text{sc}}^{(m)}(\mathbf{x},\mathbf{\hat{x}}) &= \frac{\||\operatorname{STFT}^{(m)}(\mathbf{x})|-|\operatorname{STFT}^{(m)}(\mathbf{\hat{x}})|\|_{F}}{\||\operatorname{STFT}^{(m)}(\mathbf{x})|\|_{F}},\\
    \mathcal{L}_{\text{mag}}^{(m)}(\mathbf{x},\mathbf{\hat{x}}) &= \frac{1}{N} \| \log |\operatorname{STFT}^{(m)}(\mathbf{x})| \notag\\&- \log |\operatorname{STFT}^{(m)}(\mathbf{\hat{x}})| \|_1.
\end{align}
Here, $||\cdot||$ is Frobenius norm.
Besides vocoding task, this loss is helpful for defense against adversarial attacks as well~\cite{joshi2022defense}.

\textbf{Feature Matching}~\cite{kumar2019melgan} and \textbf{Auxiliary Feature Matching}:
FM loss measures differences in activations produced by inputs $\mathbf{x}$ and $\mathbf{x'}$ in the hidden space of the discriminator i.e.
\begin{equation}
    \mathcal{L}_{\text{FM}} = \sum_{i=1}^{L} ||D_{l_i}(\mathbf{x}) - D_{l_i}(\mathbf{x'})||_1.
\end{equation}
Here, $D_{l_i}$ is the output of layer $i$.
In auxiliary FM, an external model $A$ replaces $D$, which is pre-trained on a relevant task, in our case, speaker classification.
Thus, AFM loss becomes identical to deep feature loss~\cite{johnson2016perceptual}.

\subsection{GAN adversarial losses}
\label{sec:adv_losses}
\textbf{Non-saturating loss}~\cite{goodfellow2014generative}:
The binary cross-entropy (log loss) based minimax objective for GAN is
\begin{equation}
    \min_{\mathcal{G}_{A\rightarrow B}} \max_{\mathcal{D}_{B}} \mathbb{E}_{\mathbf{a},\mathbf{b} \sim p_{A,B}}
    [\log{(\mathcal{D}(\mathbf{b}))} + \log{(1-\mathcal{D}(\mathcal{G}(\mathbf{a}))}].
\end{equation}
Authors discuss that it is easy for the discriminator to distinguish between real and fake during the beginning of training, which leads to saturation of the second term.
Hence, they proposed to maximize $log{(D(G(a))}$ for generator loss instead, which provides better gradients for the learning of the generator.

\textbf{Least Squares GAN}~\cite{mao2017least}:
Here, the authors addressed the vanishing gradient problem of regular GANs, which use sigmoid activation and log loss.
They proposed using regression loss for classification tasks and showed that this improves training stability.
The formulation is
\begin{equation}
    \min_{\mathcal{G}_{A\rightarrow B}} \max_{\mathcal{D}_{B}} \mathbb{E}_{\mathbf{a},\mathbf{b} \sim p_{A,B}}
    [(\mathcal{D}(\mathbf{b}))^2 + (1 - \mathcal{D}(\mathcal{G}(\mathbf{a}))^2
    ].
\end{equation}

\textbf{Hinge loss}~\cite{lim2017geometric}:
The authors here take a geometric view of adversarial learning.
Via hinge loss, they analyze generator and discriminator updates as learning a hyperplane to maximize the Support Vector Machine (SVM) margin
\begin{align}
    \max_{\mathcal{D}_{B}}
    \mathbb{E}_{\mathbf{b}\sim p_B}[
    \min{(0, -1+\mathcal{D}(\mathbf{b}))}] + \\ \notag
    \mathbb{E}_{\mathbf{a}\sim p_A}[
    \min{(0, -1-\mathcal{D}(\mathcal{G}(\mathbf{a})))}],
    \\
    \min_{\mathcal{G}_{A\rightarrow B}}
    -\mathbb{E}_{\mathbf{a}\sim p_A}[
    \mathcal{D}(\mathcal{G}(\mathbf{a}))].
\end{align}

\textbf{Wasserstein loss (WGAN)}~\cite{arjovsky2017wasserstein}:
This addresses the weaknesses of the original GAN in three respects: mode collapse, vanishing gradient, and hard-to-achieve Nash equilibrium.
By formulating based on Wasserstein loss (Earth Mover distance) instead of the usual Kullback-Leibler (KL) divergence and Jensen-Shannon (JS) divergence, we appropriately handle support mismatch between distributions.
The formulation is
\begin{equation}
    \min_{\mathcal{G}_{A\rightarrow B}} \max_{\mathcal{D}_{B}}
    \mathbb{E}_{\mathbf{a},\mathbf{b} \sim p_{A,B}}
    [\mathcal{D}(\textbf{b}) -
    \mathcal{D}(\mathcal{G}(\textbf{a})))
    ].
\end{equation}
We employ Wasserstein GAN-Gradient Penalty (WGAN-GP)~\cite{gulrajani2017improved} where the \emph{critic} (discriminator) gradient magnitudes are further Lipshitz-constrained to an upper value (of 0.001).

\textbf{Dual Contrastive Loss}~\cite{yu2021dual}:
Improving upon the previous GANs, here, the adversarial loss intertwines real and fake samples via the following contrastive formulation:
\begin{align}
    &\min_{\mathcal{G}_{A\rightarrow B}} \max_{\mathcal{D}_{B}} \mathbb{E}_{\mathbf{b}\sim p_{B}}[\log (1 + \sum_{\mathbf{a}\sim p_{A}} \exp({D(G(\mathbf{a})) - D(\mathbf{b})}))]
    \notag\\
    &+\mathbb{E}_{\mathbf{a}\sim p_{A}}[\log (1 + \sum_{\mathbf{b}\sim p_{B}} \exp({D(G(\mathbf{a})) - D(\mathbf{b})}))].
\end{align}

\begin{figure}[htbp]
\vspace{-4mm}
    \centering
    \includegraphics[scale=0.25]{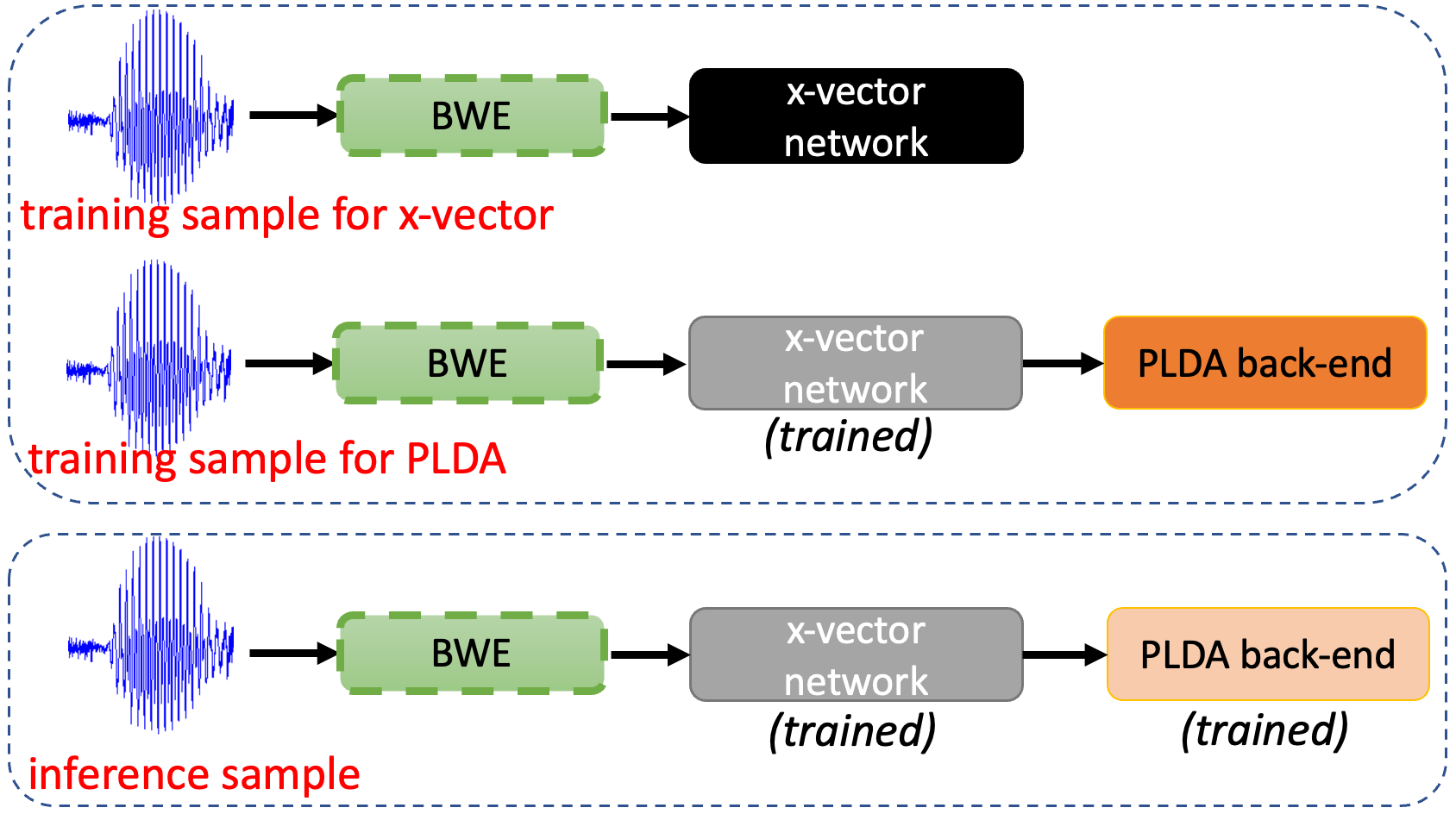}
    \caption{Three locations for deploying BWE.
    Two in training -- before x-vector network and PLDA respectively.
    One in inference -- before x-vector network.
    Corresponding results are in Table.~\ref{tab:detail}.
    }
    \label{fig:schemes}
\vspace{-4mm}
\end{figure}

\section{Proposed Evaluation Schemes}
\label{sec:prop}
In this section, we describe the methodology for the design, deployment, and evaluation of our bandwidth extension systems.
First, we describe our downstream task: Automatic Speaker Verification.
Second, we explain our greedy approach for discovering high-performing generative models.
We also discuss the possibilities of its deployment during the training and inference of the downstream task.
Finally, we analyze the effect of extension on ASV scores, perceptual quality of signals, and speaker embeddings.

\subsection{Automatic Speaker Verification}
\label{sec:asv}
The goal of ASV is to determine if two recordings contain the same dominant speaker.
One instance of such determination -- using \emph{enroll} and \emph{test} utterance -- is called a \emph{trial}.
Trials in which speakers are same are called \emph{target trials}, while for different speakers, they are called \emph{non-target trials}.
In state-of-the-art ASV systems, there is a x-vector (speaker classification) network~\cite{snyder2018x} based front-end and Probabilistic Linear Discriminant Analysis~\cite{villalba2019state} based back-end.
The former extracts speaker embeddings while the latter computes log-likelihood scores to test the binary hypothesis of ASV.
Common reporting metrics are Equal Error Rate (EER) (in \%) and minimum Decision Cost Function (DCF).
EER occurs at the decision threshold where the False Alarm (FA) and False Reject (FR) errors are equal.
Decision cost function is defined as
\begin{equation}
    C = (1 - p_\text{tar}) \times C_{\text{FA}} \times P_\text{FA} + p_\text{tar} \times C_{\text{FR}} \times P_\text{FR}.
\end{equation}
$p_\text{tar}$ is the prior probability for target speaker while $C_{\text{FA}}$ and $C_{\text{FR}}$ (set to 1 in our case) are the pre-defined costs for the two types of error.
The minimum value achieved by varying the decision threshold is called minDCF.%

\subsection{Designing time-domain GAN for bandwidth extension}
\label{sec:design}
Our primary choice for BWE model is GANs.
Since the design space of GANs can be too broad, we propose limiting the search scope.
We explore only three aspects of GAN: discriminator architecture, supervision loss function, and adversarial loss function (Sec.~\ref{sec:disc_archs} - Sec.~\ref{sec:adv_losses}).
Another reason for this exploration is that GAN performance is susceptible to design choice.
It is critical to discover a robust model suitable for further analysis.
We fix the generator architecture and other training hyper-parameters for all models.
To compare models, we note their performance on three ASV test sets.
Finally, to obtain the best GAN, we use a greedy approach by combining best discovered discriminator architecture, supervision loss function, and adversarial loss function.

\subsection{Comparison of supervised and unsupervised GANs}
\label{sec:downeval}
For fairness, we compare GAN models with simple deep regression and \emph{no extension} baselines.
Also, since there is an identity loss in CycleGAN, we propose to explore introducing such terms in deep regression and CGAN too.
Per Sec.~\ref{sec:cyclegan}, CycleGAN uses unpaired training data, but we have access to paired data due to data creation by simulation (Sec.~\ref{sec:data_descr}).
Hence, to test if CycleGAN can leverage paired data unknowingly, we also train a CycleGAN model with paired data (See \emph{paired CycleGAN} in Sec.~\ref{sec:cyclegan}).
Fig.~\ref{fig:schemes} illustrates how a trained BWE model can be utilized during the training and testing phases of downstream task.
During training, x-vector and PLDA can independently utilize BWE pre-processing.
During inference, input test signals can be extended before feeding to the trained ASV pipeline.
We apply BWE during testing \emph{blindly}, i.e., extend wideband signals too which ideally need not be extended.
Such signals are only present partially in one test set and ideally do not require an extension.
Nevertheless, we extend them, expecting our models to simulate identity operation.
In contrast with this \emph{blind} approach, we experiment not processing such signals as well.
In addition, we experiment with the Low-Frequency Replacement (LFR) technique, where the lower band of predicted signal is replaced by the lower band of the original signal.

\begin{table}[htbp]
\centering
\caption{\label{tab:discarch}
Different choice of discriminator architecture in Conditional GAN.
\textit{MS} and \textit{MP} stands for multi-scale and multi-period.
Models denoted by * in Table~I-III are identical.
}
\resizebox{0.48\textwidth}{!}{
\begingroup
\setlength{\tabcolsep}{2pt}
\begin{tabular}{@{}lccc@{}}
\toprule
 \textbf{Disc. arch.}  & \textbf{SRE16-YUE-eval40} & \textbf{SRE-CTS-superset-dev} & \textbf{SRE21-audio-eval} \\
\hline
No BWE & 7.46 / 0.382 & 5.42 / 0.217 & 18.52 / 0.662 \\
\hline
    PWG (*) & 7.88 / 0.413 & 5.21 / 0.210 & 18.03 / 0.656 \\ %
    MelGAN & 7.73 / 0.404 & 5.25 / 0.211 & 16.99 / 0.640 \\   %
    MelGAN-\textit{MS} & 7.84 / 0.406 & 5.18 / 0.210 & 16.67 / \textbf{0.636} \\ %
    HiFiGAN-\textit{MP} & 7.66 / 0.399 & 5.10 / 0.208 & \textbf{16.43} / 0.640 \\ %
    HiFiGAN-\textit{MS} & 7.81 / 0.408 & 5.11 / 0.210 & 17.16 / 0.643 \\ %
    HiFiGAN-\textit{MSMP} & 7.92 / 0.406 & 5.12 / 0.207 & 16.83 / 0.639 \\ %
    StyleMelGAN & \textbf{7.09} / \textbf{0.389} & \textbf{4.85} / \textbf{0.204} & 18.91 / 0.678 \\ %
\bottomrule
\end{tabular}
\endgroup
}
\vspace{-3mm}
\end{table}

\begin{table}[htbp]
\centering
\caption{\label{tab:suploss}
Different choice of supervision loss in Conditional GAN
}
\resizebox{0.48\textwidth}{!}{
\begingroup
\setlength{\tabcolsep}{2pt}
\begin{tabular}{@{}lccc@{}}
\toprule
\textbf{Sup. loss}  & \textbf{SRE16-YUE-eval40} & \textbf{SRE-CTS-superset-dev} & \textbf{SRE21-audio-eval} \\
\hline
No BWE & 7.46 / 0.382 & 5.42 / 0.217 & 18.52 / 0.662 \\
\hline
   MAE (*) & 7.88 / 0.413 & 5.21 / 0.210 & 18.03 / 0.656 \\ %
    MSE & 6.95 / 0.370 & \textbf{4.91} / 0.205 & \textbf{16.64} / 0.646 \\ %
    MRSTFT & 7.10 / 0.387 & 4.97 / 0.203 & 17.43 / 0.652 \\ %
    FM & 7.86 / 0.405 & 5.24 / 0.211 & 17.50 / 0.645 \\ %
    AFM & \textbf{6.89} / \textbf{0.382} & 5.38 / 0.220 & 18.40 / 0.690 \\ %
\bottomrule
\end{tabular}
\endgroup
}
\vspace{-3mm}
\end{table}

\begin{table}[htbp]
\centering
\caption{\label{tab:discloss}
Different choice of adversarial loss in Conditional GAN}
\resizebox{0.48\textwidth}{!}{
\begingroup
\setlength{\tabcolsep}{2pt}
\begin{tabular}{@{}lccc@{}}
\toprule
 \textbf{Disc. loss}  & \textbf{SRE16-YUE-eval40} & \textbf{SRE-CTS-superset-dev} & \textbf{SRE21-audio-eval} \\
\hline
No BWE & 7.46 / 0.382 & 5.42 / 0.217 & 18.52 / 0.662 \\
\hline
   LSGAN (*) & 7.88 / 0.413 & 5.21 / 0.210 & 18.03 / 0.656 \\ %
   Non-saturating & \textbf{7.11} / \textbf{0.382} & \textbf{4.85} / \textbf{0.202} & \textbf{17.29} / 0.652 \\ %
   Hinge & 7.45 / 0.399 & 5.21 / 0.211 & 17.64 / 0.652 \\ %
   Wasserstein & 7.44 / 0.387 & 5.06 / 0.205 & 17.44 / \textbf{0.651} \\ %
   DCL & 7.20 / 0.388 & 5.18 / 0.210 & 17.71 / 0.661 \\ %
\bottomrule
\end{tabular}
\endgroup
}
\vspace{-2mm}
\end{table}

\subsection{Per-condition score analysis}
We are also interested in analyzing per-condition results since auxiliary condition information is available for our two test sets.
In the context of BWE, two types of speech are of utmost interest: CTS (conversational telephone narrowband speech) and AFV (Audio from Video wideband speech \emph{in the wild}).
Therefore, we focus on three trial types: CTS-CTS, CTS-AFV, and AFV-AFV.
We also study trials based on language, gender, and recording device.
However, averaged scores do not convey all information about distribution shifts in verification scores.
We propose to study this in two ways.
One, plot score histograms of CTS-AFV trials before and after BWE.
Two, plot CTS-CTS, CTS-AFV, and AFV-AFV scores and examine if BWE brings them closer.

\vspace{-2mm}
\subsection{Perceptual quality and speaker embedding analysis}
\label{sec:perceptsne}
To further analyze our BWE systems, we explore the effects of extension on perceptual quality and speaker embeddings of signals.
In the absence of explicit perceptual quality objectives in training, it is paramount to explore if there is a correlation between the downstream performance and the output speech quality.
For perceptual and intelligibility evaluation, we mainly choose \emph{full reference} automated measures like PESQ~\cite{rix2001perceptual} and Extended STOI (ESTOI)~\cite{jensen2016algorithm} respectively.
PESQ is a psychoacoustics-based measure to estimate the quality of speech affected by perceptual distortions while adjusting for time lags and loudness mismatch.
ESTOI improves upon STOI in measuring intelligibility in the distorted speech by considering correlations between frequency bands.
By doing this analysis for deep regression, CGAN, and CycleGAN models, we can also quantify the effect of using paired training data and generative modeling.
To analyze embeddings, we propose to utilize t-SNE~\cite{van2008visualizing}, a popular non-linear dimension reduction and visualization technique.
Our previous work~\cite{kataria2021deep} showed that the effect of domain adaptation via generative models is apparent as a shift in the t-SNE plot.
Thus, we propose to visualize narrowband train and test data embeddings before and after extension. %
We also propose visualizing how CTS and AFV audio from the same speaker cluster before and after the extension.

\section{Experimental Setup}
\label{sec:exp}
\subsection{Data Description}
\label{sec:data_descr}
All signals in our experiments follow the sampling frequency of 16000 samples per second and amplitude normalized to [-1,1].
We know that the maximum frequency information ($f_{\text{max}}$) in a signal depends on its original sampling frequency ($f_s$) and equals to $f_s/2$.
We refer to signals with $f_{\text{max}}$ of 4KHz and 8KHz as narrowband and wideband, respectively.
A wideband signal, thus, must have a sampling frequency of at least 16KHz as per the Nyquist theorem.
As baseline, we resample the narrowband signals in train and test sets from their original sampling frequency of 8KHz to 16KHz using linear upsampling.%

We have three types of data available during training: \emph{narrow}, \emph{wide}, and \emph{wide\_down}.
\emph{narrow} data refers to narrowband telephone corpus from SRE Superset~\cite{sadjadi2021nist} and SRE16 eval data~\cite{reynolds20172016,villalbajhu} which includes Tagalog and Cantonese (YUE) languages.
\emph{wide} data refers to wideband VoxCeleb~\cite{nagrani2020voxceleb} -- also called as \emph{microphone} data.
VoxCeleb (1\&2 combined) contains 2700+ hrs of audio from 7365 speakers in the wild.
\emph{wide\_down} data refers to narrowband VoxCeleb, created by downsampling and subsequent linear upsampling of \emph{wide} data, making \emph{wide\_down} a narrowband dataset as well.
We evaluate ASV on three test sets.
(1) \emph{SRE16-YUE-eval40}~\cite{villalbajhu,reynolds20172016} (40\% speakers (40) from evaluation set of SRE16 Cantonese).
(2) \emph{SRE-CTS-superset-dev}~\cite{villalba22b_odyssey}.
This contains 99 speakers from CMN (Mandarin) and YUE (Cantonese) languages.
It is balanced between genders and contains 22M trials.
(3) \emph{SRE21-audio-eval}~\cite{sadjadi20222021} (SRE21 eval set).
It contains 6M trials and various languages, channels, devices, et cetera.
This test set, contrary to others, also consists of wideband AFV signals.

\subsection{Bandwidth Extension training}
For training bandwidth extension models, we use only \emph{wide\_down} and \emph{wide} data.
As explained in Sec.~\ref{sec:bwe_models}, supervised models (deep regression, CGAN) use paired samples while for unsupervised models (CycleGAN), we use unpaired samples.
We also found that silence regions are critical for BWE training, hence we preserve them in training samples of BWE.

\subsubsection{Deep regression training}
We train Conv-TasNet
with 4~s audio segments with \emph{wide\_down} input and corresponding \emph{wide} target.
Batch size is 128, number of epochs are 70 (defined as 100~h of speech), optimizer is Adam~\cite{kingma2014adam} with betas=(0.9, 0.999), and objective function is temporal MAE loss.
The learning rate of 0.0005 decreases by half when validation loss does not decrease by at least 1\% for three epochs.

\subsubsection{CGAN training}
Here, we train using Alternating Gradient Descent (AGD)~\cite{goodfellow2014generative}.
One training step consists of one update of discriminator (given fixed generator) and two updates of generator (given fixed discriminator).
Due to the sensitivity of GAN training with Conv-TasNet architecture, lower precision training is disabled.
$\lambda_{\text{sup}}=0.1$. %
Learning rates for generator and discriminator are 0.0002 and 0.0001, respectively, which decrease linearly with every training step until a minimum value of 1e-07.
The sequence length for training is 3~s, the batch size is 16, the number of epochs is 15 (defined as 50 h of speech), and Adam betas are (0.5,0.999).

\subsubsection{CycleGAN training}
Here, the maximum value of learning rates for the generator and discriminator are 0.0004 and 0.0002, respectively, and the minimum value is 1e-08.
For the first two epochs, the learning rate linearly increases from minimum to maximum value.
The learning rate is constant at the maximum values for the subsequent three epochs.
The learning rate decreases to a minimum value following a cosine function for the final ten epochs.
The batch size is 8, and all other training details are identical to CGAN.

\subsection{Baseline speaker verification system}
\label{sec:asv_details}
This work follows our submission to the fixed-condition NIST Speaker Recognition Challenge (SRE) 2021 challenge~\cite{sadjadi2021nist}~\footnote{\url{https://github.com/hyperion-ml/hyperion/tree/master/egs/sre21-av-a}}.
The x-vector model used is Light-ResNet~\cite{villalba2020advances} with speaker embedding dimension of 256.
Loss function is Additive Angular Margin (AAM) loss with margin $m=0.3$.
Input features are 80-D Log-Mel FilterBank (LMFB), created on-the-fly.
For baseline, training uses \emph{wide}, \emph{wide\_down}, and unmodified\emph{narrow} (except for linear upsampling of narrowband data to 16 kHz).
Noise and reverberation augmentation is done on-the-fly using MUSAN~\cite{snyder2015musan} and Aachen Impulse Response (AIR) Database corpus~\footnote{\url{http://www.openslr.org/resources/28}}.
After 200-D Linear Discriminant Analysis (LDA) pre-processing on 256-D embeddings, 150-D PLDA is used.
We remove silence frames from the x-vector and PLDA training using a simple energy-based voice activity detector.

\begin{table*}[htbp]
\centering
\caption{\label{tab:comp}
Comparison of deep regression, CGAN, and CycleGAN}
\begin{tabular}{@{}lccccc@{}}
\toprule
\textbf{BWE system} & \textbf{Training data paired/unpaired} & \textbf{$\lambda_{\text{id}}$} & \textbf{SRE16-YUE-eval40} & \textbf{SRE-CTS-superset-dev} & \textbf{SRE21-audio-eval} \\
\hline
- & - & - & 7.46 / 0.382 & 5.42 / 0.217 & 18.52 / 0.662 \\ %
\hline
Mapping & paired & 0 & 7.28 / 0.395 & 4.99 / 0.211 & 16.72 / 0.642 \\ %
Mapping & paired & 0.5 & 7.60 / 0.388 & 5.21 / 0.210 & 18.34 / 0.664 \\ %
CGAN & paired & 0 & 7.07 / 0.378 & 5.05 / 0.207 & 15.99 / 0.630 \\ %
CGAN & paired & 0.1 & 7.22 / 0.378 & 5.04 / 0.206 & \textbf{15.95} / \textbf{0.629} \\ %
CycleGAN & unpaired & 10 & \textbf{6.86} / \textbf{0.371} & \textbf{4.97} / \textbf{0.205} & 17.27 / 0.656 \\ %
CycleGAN\_unsupervised & paired & 10 & 7.10 / 0.377 & 4.96 / 0.204 & 19.49 / 0.674 \\ %
CycleGAN\_supervised & paired & 10 & 7.71 / 0.400 & 5.27 / 0.211 & 18.69 / 0.667 \\ %
\bottomrule
\end{tabular}
\vspace{-4mm}
\end{table*}

\begin{table*}[htbp]
\centering
\caption{\label{tab:detail}
Bandwidth Extension of x-vector and PLDA training data.
Test set is extended per blind strategy. For x-vector training, here ``original data'' is \{\emph{wide}, \emph{wide\_down}, \emph{narrow}\}, while ``extended data'' is \{\emph{wide}, $M$(\emph{wide\_down}), $M$(\emph{narrow})\}. $M(\cdot)$ is the extension model.
}
\begin{tabular}{@{}lcccc@{}}
\toprule
\textbf{PLDA data} & \textbf{PLDA data extended} & \textbf{SRE16-YUE-eval40} & \textbf{SRE-CTS-superset-dev} & \textbf{SRE21-audio-eval} \\
\hline
\multicolumn{3}{@{}l}{\textbf{\underline{x-vector trained and fine-tuned on original data}}}  && \\
\emph{wide}, \emph{wide\_down}, \emph{narrow} & - & 6.83 / 0.359 & 4.71 / 0.202 & 15.93 / 0.623 \\ %
\emph{wide}, \emph{wide\_down}, \emph{narrow} & \emph{narrow} & 6.39 / 0.352 & 4.91 / 0.204 & 14.82 / 0.599 \\ %
\emph{wide}, \emph{narrow} & \emph{narrow} & \textbf{5.27} / \textbf{0.307} & \textbf{4.01} / \textbf{0.174} & \textbf{14.33} / \textbf{0.591} \\ %
\hline
\multicolumn{3}{@{}l}{\textbf{\underline{x-vector trained on original data and fine-tuned on extended data}}} && \\
\emph{wide}, \emph{wide\_down}, \emph{narrow} & - & 6.88 / 0.372 & 5.33 / 0.215 & 15.31 / 0.614 \\ %
\emph{wide}, \emph{wide\_down}, \emph{narrow} & \emph{narrow} & 6.83 / 0.373 & 5.17 / 0.210 & 15.59 / 0.615 \\ %
\emph{wide}, \emph{narrow} & \emph{narrow} & \textbf{5.43} / \textbf{0.315} & \textbf{4.11} / \textbf{0.175} & \textbf{14.88} / \textbf{0.597} \\ %
\hline
\multicolumn{3}{@{}l}{\textbf{\underline{x-vector trained and fine-tuned on extended data}}} && \\
\emph{wide}, \emph{wide\_down}, \emph{narrow} & - & 7.64 / 0.436 & 5.53 / 0.220 & 16.32 / 0.650 \\ %
\emph{wide}, \emph{wide\_down}, \emph{narrow} & \emph{narrow} & 7.20 / 0.413 & 5.35 / 0.219 & 16.46 / 0.644 \\ %
\emph{wide}, \emph{narrow} & \emph{narrow} & \textbf{5.67} / \textbf{0.363} & \textbf{4.16} / \textbf{0.187} & \textbf{15.29} / \textbf{0.616} \\ %
\bottomrule
\end{tabular}
\vspace{-4mm}
\end{table*}

\section{Results}
\label{sec:results}
\subsection{Designing Conditional GAN based Bandwidth Extension}
\label{sec:res_cgan}
Here, we present the results for exploring discriminator architecture, supervision loss, and adversarial loss for CGAN in Tables \ref{tab:discarch}, \ref{tab:suploss}, and \ref{tab:discloss} respectively.
Results are in the format EER/minDCF, and we discuss in detail such exploration for CGAN only.
In Table~\ref{tab:discarch}, the first row has results without BWE.
In other rows, we expand real narrowband data (\emph{narrow}) of PLDA and test set using \emph{blind} extension strategy (Sec.~\ref{sec:downeval}).
The x-vector is trained a priori on unmodified wideband and narrowband data and used as is.
Row 2 (marked as *) is the model identical to row 2 of the other two tables.
We see that multi-period (\textit{MP}) and multi-scale (\textit{MS}) versions of discriminator architecture bring significant benefits in \emph{SRE21-audio-eval}, partly due to more parameters.
StyleMelGAN performs best except on \emph{SRE21-audio-eval}.

Consider Table~\ref{tab:suploss}.
Note that no choice of supervision loss function gives the best results on all test sets.
Also, we found that the combination of several losses is ineffective.
Several loss functions improve w.r.t. \emph{no extension} baseline.
MSE loss gives the best performance overall.
MAE is worse than MSE.
However, it was the preferred choice in our previous works~\cite{kataria2021deep,kataria2022joint}.
AFM loss gives the best performance on \emph{SRE16-YUE-eval40}, but this comes with the cost of increased computation due to the usage of the auxiliary model.
We use the outputs of five ResNet blocks of our x-vector network for computing AFM loss.

Now consider Table~\ref{tab:discloss}.
Here, surprisingly, the best performance is achieved by the simplest loss, i.e., non-saturating loss (from original GAN work~\cite{goodfellow2014generative}) -- contrary to our preferred choice of LSGAN loss in previous works~\cite{kataria2021deep}.
Wasserstein loss also gives promising results at the cost of tuning an additional hyper-parameter of gradient clipping factor.

Finally, we train a CGAN with the best attributes from the above experiments, i.e., with MSE supervision loss, non-saturating adversarial loss, and StyleMelGAN discriminator architecture.
We perform similar exploration experiments for CycleGAN and find the best attributes to be MAE supervision (i.e., cycle \& identity) loss, LSGAN adversarial loss, and HiFiGANMultiPeriod discriminator architecture.
This individual in-depth exploration of CGAN and CycleGAN allows us to compare them fairly.
Note that \emph{SRE21-audio-eval} has significantly worse results than the other two test sets because it is a highly mismatched dataset (w.r.t. language, channel, et cetera).
A much stronger recognition system is needed to achieve better performance~\cite{avdeeva2021stc,villalba22b_odyssey}.

\subsection{Comparison of supervised and unsupervised BWE}
In Table~\ref{tab:comp}, we compare deep regression, CGAN, and CycleGAN.
Our first observation is that the most straightforward scheme, i.e., deep regression, brings significant improvements in all test sets, particularly in EER.
However, its performance on \emph{SRE16-YUE-eval40} is lacking.
Adding identity loss makes it significantly worse.
We train CGAN with the best attributes discovered in the previous section.
It gives a strong performance, but the utility of identity loss is inconclusive here.
CycleGAN gives the best results on two test sets, as shown in the boldface.
Identity loss is crucial for CycleGAN training, so we do not experiment with removing it.
``CycleGAN\_unsupervised paired'' refers to vanilla CycleGAN but trained with paired data.
This model cannot leverage pairing information inherently showing that paired data is detrimental to CycleGAN.
``CycleGAN\_supervised paired'' is the CycleGAN model where two supervision losses for both directions are added to the formulation since we know the expected output (from paired data).
This model is thus comparable to using two tied CGANs and turns out to be better than ``CycleGAN\_unsupervised paired'' but on only \emph{SRE21-audio-eval}.
Our best models are vanilla CGAN and CycleGAN, which we use for further analysis.
We encourage the reader to listen to a few samples~\footnote{\url{https://github.com/saurabh-kataria/BWE-samples}}.
We also visually compare the three BWE techniques in Fig.~\ref{fig:spectrograms}.
We specifically highlight the quality of CycleGAN prediction here.

\vspace{-2mm}
\subsection{Effect of extending x-vector and PLDA training data}
In Table~\ref{tab:detail}, we investigate the effect of extending training data of PLDA and x-vector.
The results are divided into three parts.
Complete results are in Appendix~\ref{sec:appendix} but we discuss a subset of results here.
In the first part, the x-vector is trained (on 4~s chunks) and fine-tuned (on 10-60~s chunks) on original unextended data i.e. \emph{wide}, \emph{wide\_down}, and \emph{narrow}.
The first column lists the data used for PLDA training, while the second column denotes which one of them is extended.
We use CGAN for extension here.
The results here are better than Table~\ref{tab:comp} since the x-vector is fine-tuned on long recordings which significantly improves ASV performance~\cite{villalba2019state}.
We observe that removing synthetic data (\emph{wide\_down}) from PLDA training and not extending \emph{wide} data is a better choice (see also Appendix~\ref{sec:appendix}).
In the second part, the x-vector is trained on original data but during fine-tuning, narrowband data i.e. \emph{wide\_down} and \emph{narrow} are extended.
The observations here are similar to the previous part but with slight degradation.
In the third part, the x-vector is trained and fine-tuned on wideband (\emph{wide}) and extended narrowband data.
We observe even more degradations here.
We finally report extending x-vector data as inconclusive.
In future work, we can use larger x-vector networks like the ones used in \cite{villalba22b_odyssey}.

\vspace{-2mm}
\subsection{SRE21 results per trial condition}
In Table~\ref{tab:condition}, we note the absolute and relative improvement in EER on \emph{SRE21-audio-eval}.
We only report CGAN results here since we found similar observations for CycleGAN.
The overall improvement averaged across all conditions is significant (-13.67\%).
Mismatched condition CTS-AFV gives only -2.7~\% relative improvement.
CTS-CTS performance (-0.5~\%) is unaffected as intended.
AFV-AFV is matched trial but is adversely affected by the extension.
We address this issue in next section (Sec.~\ref{sec:testtime}).
Most of the gains come from trials other than CTS-CTS, CTS-AFV, and AFV-AFV.
We liken this more generic benefit to \emph{calibration}, that is, extension is shifting scores in order to make speaker verification robust to different acoustic environments.
Finally, we note that improvement is highest when test language is matched with train (ENG-ENG).

\begin{figure*}[htbp]
    \centering
    \subfloat[\centering
    CGAN extension
    ]{{\includegraphics[width=7.5cm]{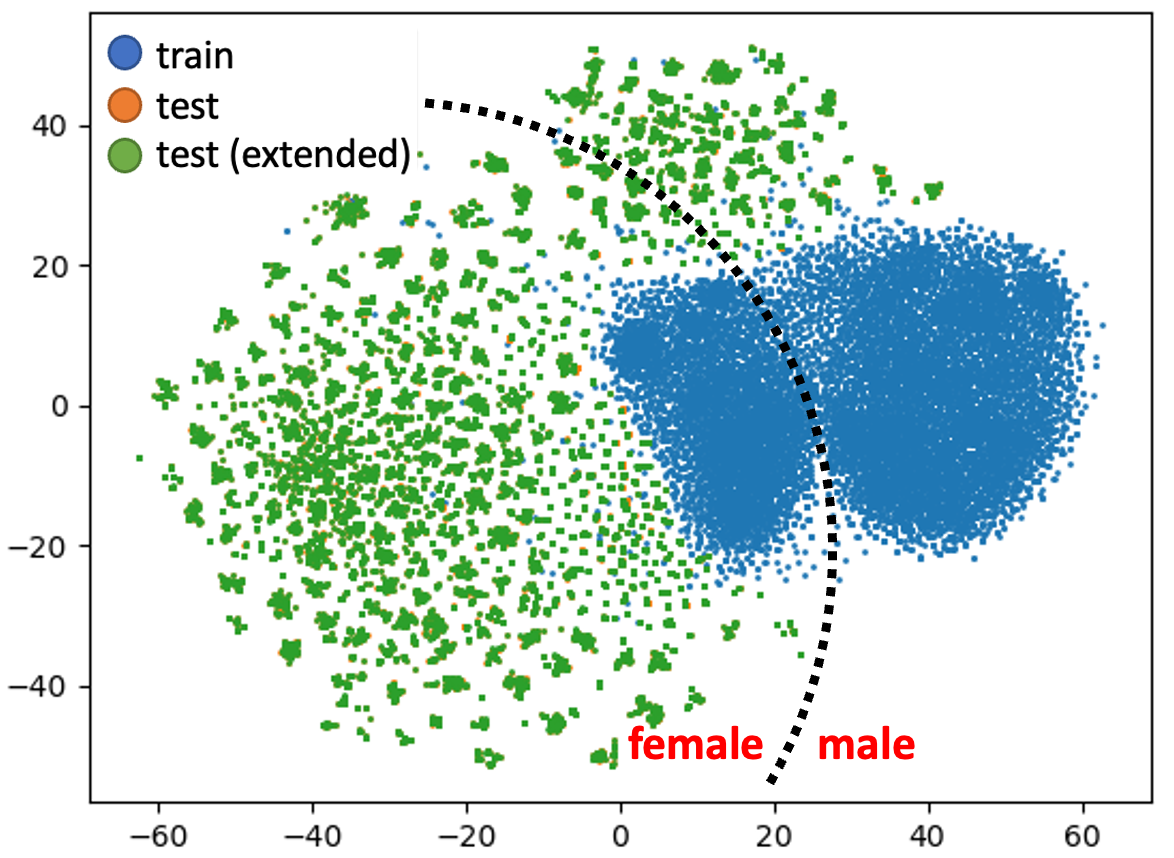}}}%
    \qquad
    \subfloat[\centering
    CycleGAN extension
    ]{{\includegraphics[width=7.5cm]{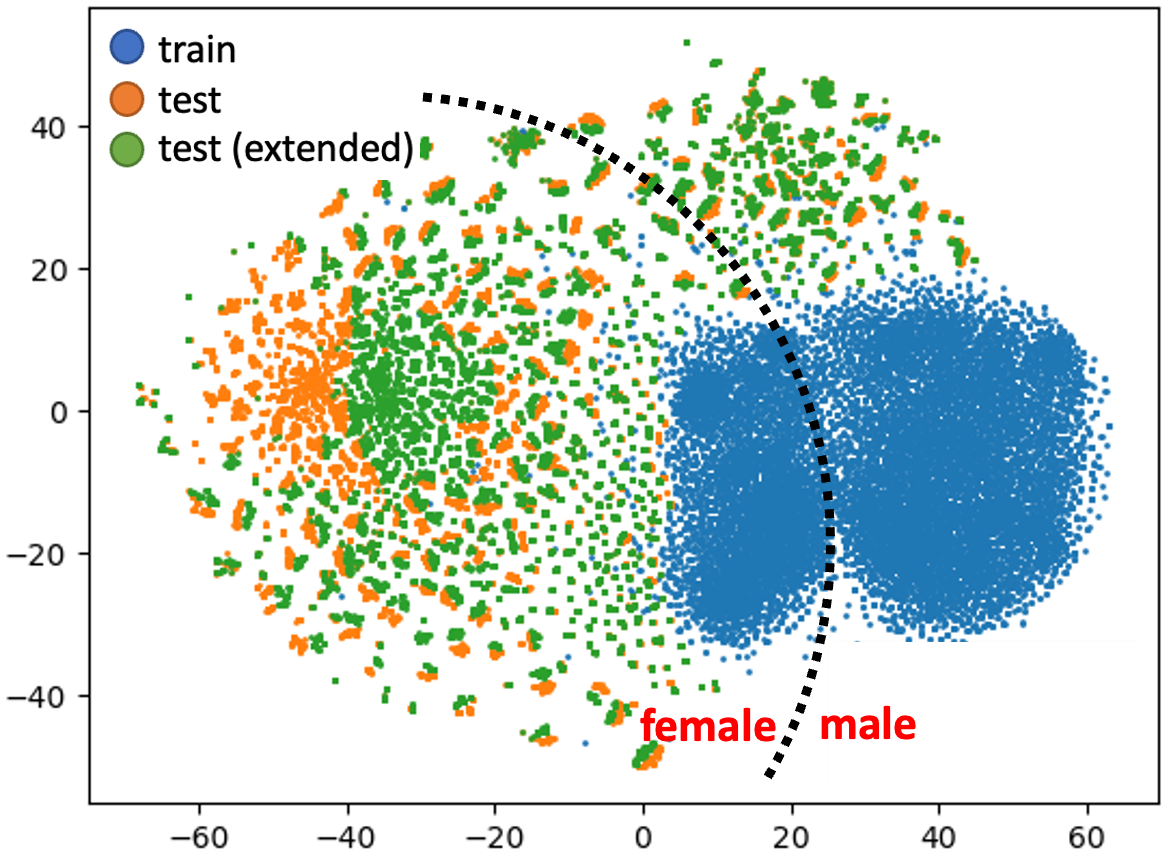}}}%
    \caption{t-SNE visualization of embeddings of wideband training data and narrowband test data (with and without extension). ``train'' refers to wideband VoxCeleb. ``test'' refers to narrowband SRE21 test set. ``test (extended)'' refers to SRE21 test set extended by CGAN or CycleGAN.
    We observe that CycleGAN causes a shift in distribution.
    For CGAN, there is no shift, as noted from the perfect overlap of orange and green dots.
    }
    \label{fig:emb_train_test}
    \vspace{-4mm}
\end{figure*}
\begin{figure}[htbp]
    \centering
    \includegraphics[width=7.25cm]{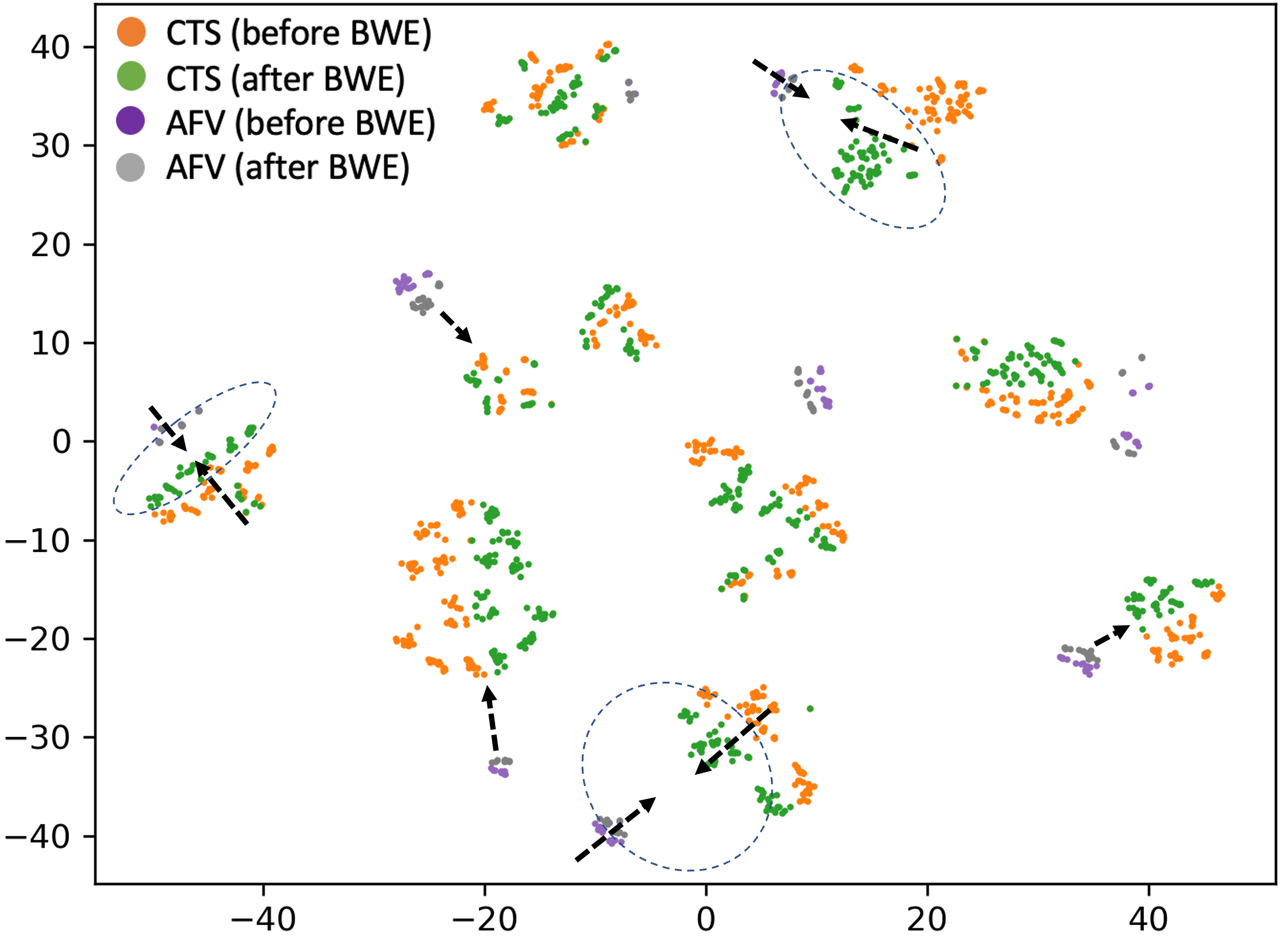} %
    \caption{t-SNE visualization of CTS and AFV recordings of 10 speakers before and after BWE. The four types of signals come closer by extension, as denoted by black arrows—dotted ellipse highlight a few cases where CTS and AFV signals come close after extension.}
    \label{fig:indivspk}
    \vspace{-4mm}
\end{figure}

\begin{figure*}[htbp]
    \centering
    \subfloat[\centering Target trials
    ]{{\includegraphics[width=7.5cm]{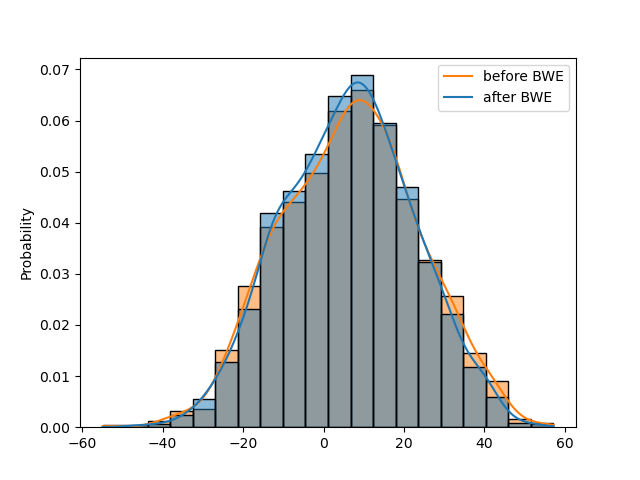} }}%
    \subfloat[\centering Non-target trials ]{{\includegraphics[width=7.5cm]{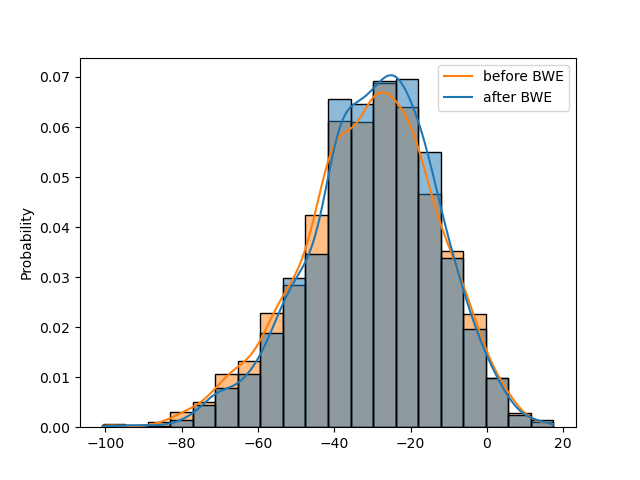} }}%
    \caption{CTS-AFV score distribution before and after CGAN extension for the SRE21 test set.
    CTS and AFV signals are both extended.
    Gray denotes the overlap of orange and blue colors.
    Blue dominating in the middle suggests extension shifts scores towards the mean.
    }
    \label{fig:score1}
    \vspace{-6mm}
\end{figure*}

\begin{figure*}[htbp]
    \centering
    \subfloat[\centering Before extension
    ]{{\includegraphics[width=6.90cm]{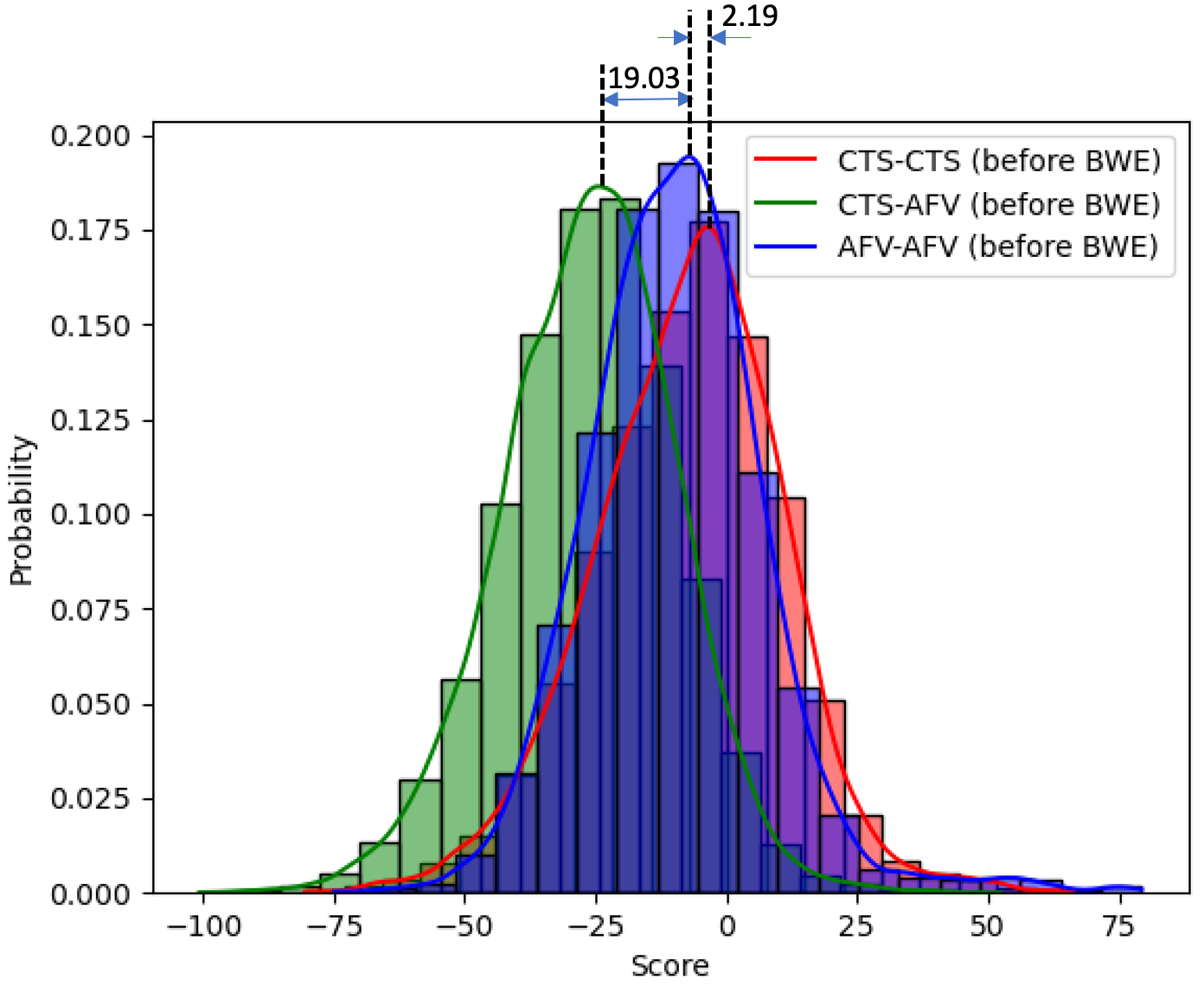}}}%
    \quad \qquad  %
    \subfloat[\centering After extension
    ]{{\includegraphics[width=6.90cm]{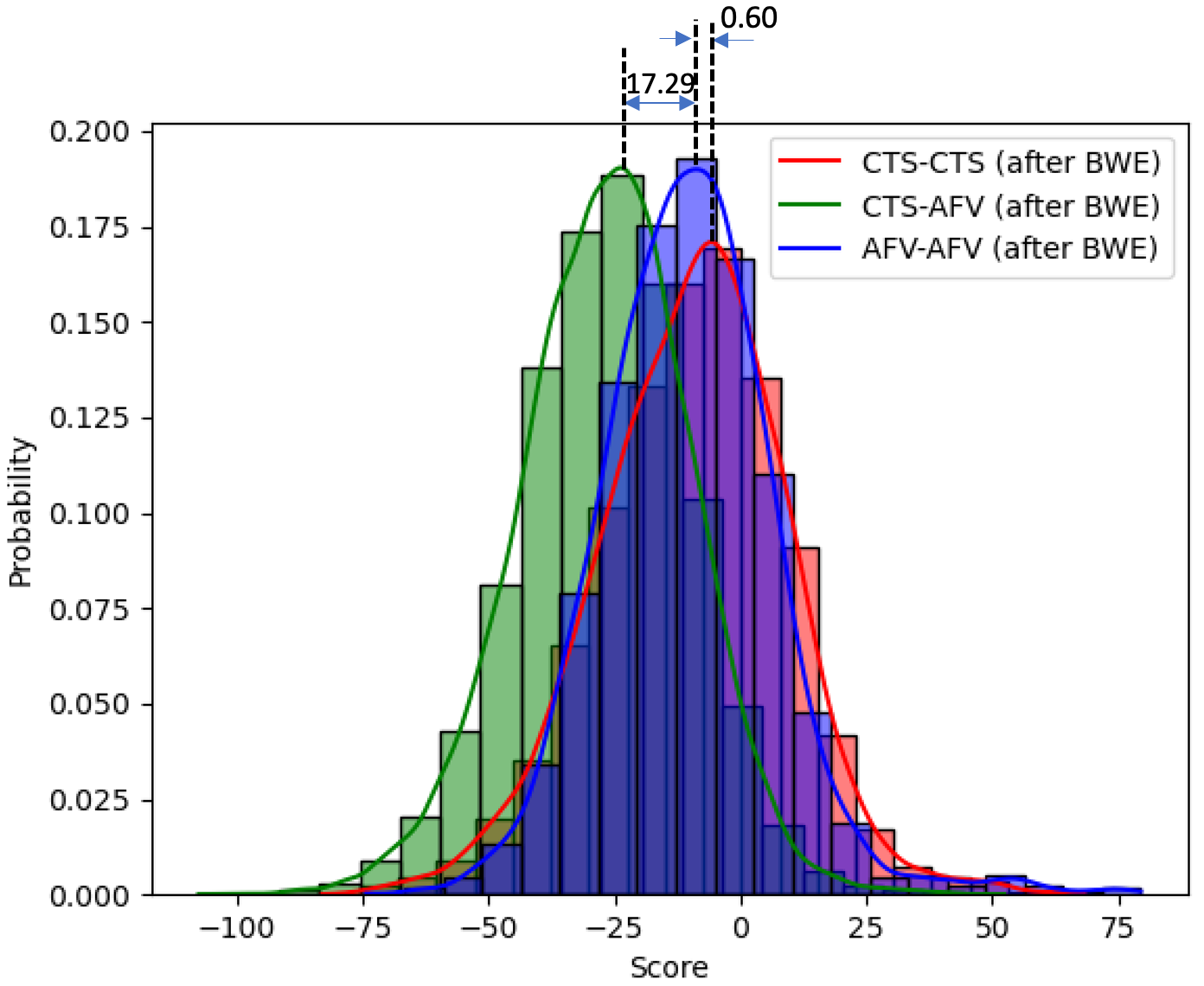}}}%
    \caption{Score distribution of CTS-CTS, CTS-AFV, and AFV-AFV trials before and after CGAN extension. BWE brings the peaks of all curves closer.}
    \label{fig:score2}
    \vspace{-5mm}
\end{figure*}

\begin{table}[htbp]
\centering
\begingroup
\setlength{\tabcolsep}{3pt}
\caption{\label{tab:condition}
EER improvement by trial condition.
}
\begin{tabular}{@{}lcHcHc@{}}
\toprule
 \textbf{Condition} & \textbf{No BWE} & \textbf{Deep regression} & \textbf{CGAN} & \textbf{CycleGAN} & \textbf{\% change} \\ %
\hline
   Overall & 18.52 & 16.72 / 0.642 & \textbf{15.99} & 17.27 / 0.656 & -13.67\%\\
\hline
   CTS-CTS & 8.85 & 9.41 / 0.474 & \textbf{8.81} & 8.99 / 0.477 & -0.5\% \\
   CTS-AFV & 16.69 & 16.55 / 0.700 & \textbf{16.24} & 17.05 / 0.710 & -2.7\% \\
   AFV-AFV & \textbf{2.89} & 3.84 / 0.184 & 3.49 & 3.89 / 0.192 & +20.8\%\\
\hline
   ENG-ENG & 17.98 & 14.70 / 0.605 & \textbf{14.02} & 16.17 / 0.625 & -22.0\% \\
   ENG-YUE & 18.83 & 16.85 / 0.625 & \textbf{15.93} & 17.33 / 0.625 & -15.4\% \\
   YUE-YUE & 19.38 & 17.55 / 0.707 & \textbf{16.76} & 18.09 / 0.727 & -13.5\% \\
\hline
   Same device & \textbf{4.90} & 5.58 / 0.336 & 5.24 & 5.56 / 0.345 & +6.9\% \\
   Different device & 25.53 & 23.10 / 0.910 & \textbf{22.11} & 23.93 / 0.927 & -13.4\% \\
\bottomrule
\end{tabular}
\endgroup
\vspace{-4mm}
\end{table}

\begin{table}[htbp]
\centering
\caption{\label{tab:schemes}
Comparing different test-time BWE schemes.
}
\resizebox{0.48\textwidth}{!}{
\begingroup
\setlength{\tabcolsep}{3pt}
\begin{tabular}{@{}lccccc@{}}
\toprule
 \textbf{Condition} & \textbf{No BWE} & \textbf{expand\_all} & \textbf{expand\_narrow} & \textbf{LFR} & \textbf{expand\_narrow} \\
 & & & & & \textbf{+ LFR} \\
\hline  %
   Overall & 18.52 & \textbf{15.99} & 17.29 & 17.29 & 17.90 \\
\hline
   CTS-CTS & 8.85 & 8.81 & 8.76 & \textbf{8.75} & 8.89 \\
   CTS-AFV & 16.69 & \textbf{16.24} & 16.80 & 16.78 & 17.03 \\
   AFV-AFV & \textbf{2.89} & 3.49 & 3.03 & 2.97 & 3.00 \\
\hline
   ENG-ENG & 17.98 & \textbf{14.02} & 15.67 & 15.71 & 16.77 \\
\bottomrule
\end{tabular}
\endgroup
}
\end{table}

\vspace{-3mm}
\subsection{Different test-time extension schemes}
\label{sec:testtime}
In Table~\ref{tab:schemes}, we report results for various test-time extension schemes on \emph{SRE21-audio-eval}.
Here, we again report results for only CGAN since we found similar observations for CycleGAN.
\emph{Blind} extension test scheme, denoted by \emph{expand\_all} in column 3, extends narrow as well as wideband test signals, which is undesirable.
\emph{Expand\_narrow} extends only narrowband signals and improves AFV-AFV performance.
It hurts other conditions and overall performance because DNN induces a mismatch between extended and unextended signals.
Low-Frequency Replacement strategy (\emph{LFR}), which does not modify lower band frequency information, does not improve overall but gives good AFV-AFV performance.
In the last row, a combination of \emph{expand\_narrow} and \emph{LFR} also does not prove beneficial.
Thus, devised schemes have worse performance than baseline.
They are significantly worse in conditions other than CTS-CTS, CTS-AFV, and AFV-AFV.
We report one such condition: ENG-ENG.

\vspace{-2mm}
\subsection{Effect of extension on speaker embeddings}
In Fig.~\ref{fig:emb_train_test}, we plot t-SNE embeddings of x-vector representations of training and test data.
The black dotted curves separate the male and female embeddings.
The model for extension used in the two figures is CGAN and CycleGAN, respectively.
Blue markers denote wideband training data (i.e., VoxCeleb or \emph{wide}).
Orange markers denote test set samples from \emph{SRE21-audio-eval}, while green markers denote their extended counterpart.
We observe the following.
First, there is a clear separation w.r.t. gender in both plots.
Second, in the case of CGAN, the extension does not achieve the noticeable shift in t-SNE space, while for CycleGAN, it is pronounced.
Third, for the female gender, the shift in embeddings is higher, which we did not observe in the other test set: \emph{SRE-CTS-superset-dev}.
This experiment reveals the difference in the extension behavior of both models.
This shift caused by CycleGAN is akin to domain adaptation effect~\cite{kataria2021deep}.
In Fig.~\ref{fig:indivspk}, we plot the t-SNE of CTS and AFV embeddings before and after CGAN extension.
We use ten speakers for this analysis and can see the four types of signals cluster per speaker.
Extension brings all types of signals (\{CTS, AFV\} x \{before extension, after extension\}) closer, as denoted by black arrows.
Dotted ellipses highlight a few cases where CTS and AFV signals come significantly closer after extension.

\vspace{-2mm}
\subsection{Effect of extension on verification score distribution}
\label{sec:ctsafvscore}
In Fig.~\ref{fig:score1}, we plot the distributions for target and non-target trials before and after CGAN extension of \emph{SRE21-audio-eval}.
Blue colors (denoting after extension) dominate the bars in the histogram's middle.
For both target and non-target trials, extension brings scores closer to the mean.
This observation holds for other trial conditions too.
We interpret this as a calibration-like effect.
Similar to motivation for Fig.~\ref{fig:indivspk}, in Fig.~\ref{fig:score2}, we demonstrate through histograms that extension brings scores of CTS-CTS, CTS-AFV, and AFV-AFV closer.
We can see a significant decrease in the difference between the mean of the three curves.

\vspace{-2mm}
\subsection{Perceptual quality and relation to downstream metric}
A previous work~\cite{siddiqui2020using} showed that the perceptual quality of DNN outputs (in our case, extended signals) does not necessarily correlate with the performance of downstream tasks.
We investigate this on our experimental setup in Table~\ref{tab:relation}.
We use PESQ, ESTOI, LSD, and MSE metrics to measure signal distortion.
We also devise \emph{deep feature MSE}, which measures the MSE error (w.r.t. ground truth) measured using activations of signals in an auxiliary network (x-vector, in our case).
In addition to the above measures, we note the relative improvements in EER and minDCF (averaged across our three test sets).
We first note that unextended signals have the highest perceptual quality and all extension methods give a low PESQ value since we do not optimize for it.
Since all extension systems use a form of supervision loss, we do not notice significant degradation in ESTOI, LSD, and time-domain MSE.
Deep regression expectedly gives the best results on these metrics.
Also, note that deep feature MSE is constant across all methods, i.e., speaker information is not lost.
It is also evident from good ASV performance with all methods.
CGAN gives the best average relative improvement in EER and minDCF.
Deep regression gives competitive results but fails to improve minDCF.
CycleGAN gives the worst perceptual quality but excellent ASV performance and spectrograms (Fig.~\ref{fig:spectrograms}).

\begin{table*}[htbp]
\centering
\begingroup
\setlength{\tabcolsep}{3pt}
\caption{\label{tab:relation}
Relation between perceptual quality and downstream performance. Metrics denoted by $\uparrow$ ($\downarrow$) are higher (lower) the better.}
\resizebox{0.96\textwidth}{!}{
\begin{tabular}{l|cc|ccc|cc|cc}
\toprule
    \textbf{BWE system}  & \textbf{PESQ} ($\uparrow$) & \textbf{ESTOI} ($\uparrow$) & \textbf{LSD} ($\downarrow$) & \textbf{time-domain MSE} ($\downarrow$) & \textbf{deep feature MSE} ($\downarrow$) & \textbf{EER} ($\downarrow$) & \textbf{$\Delta$ EER} ($\downarrow$) & \textbf{minDCF} ($\downarrow$) & \textbf{$\Delta$ minDCF} ($\downarrow$) \\
\hline
   No BWE & \textbf{3.846} & 0.989 & 1.102 & 3.233 & 913.410 & 10.47 & 0\% & 0.420 & 0\% \\
   Deep regression & 3.659 & \textbf{0.993} & \textbf{0.471} & \textbf{1.042} & 913.414 & 9.66 & -6.69\% & 0.416 & -0.79\% \\
   CGAN & 3.514 & 0.985 & 0.790 & 1.248 & \textbf{913.363} & \textbf{9.37} & \textbf{-8.57\%} & \textbf{0.405} & \textbf{-3.50\%} \\
   CycleGAN & 2.055 & 0.977 & 1.301 & 3.816 & 913.408 & 9.70 & -7.70\% & 0.410 & -3.11\%\\
\bottomrule
\end{tabular}
}
\endgroup
\vspace{-5mm}
\end{table*}

\section{Conclusions and future work}
In this work, we comprehensively evaluate bandwidth extension of narrowband data for the downstream task of telephony speaker verification.
Our ASV system is based on the state-of-the-art pipeline: x-vector front-end, PLDA back-end, data augmentation, and mixed bandwidth training data.
We focused on time-domain BWE models due to their flexible usage during the training and inference of ASV.
We discover high-performing supervised and unsupervised GANs like conditional GAN and CycleGAN respectively.
We first extensively tune conditional GAN to (1) demonstrate the sensitivity of GAN performance to design and (2) derive the best model suitable for fair comparison and further analysis.
With this tuning, we discover vastly different designs for CGAN and CycleGAN.
Comparing these best models on three real test sets, we find the deep regression baseline to be strong.
Unsupervised CycleGAN performs on par with supervised CGAN and even surpasses performance on a few metrics.
These results are however obtained via extension of test set and the real narrowband (\emph{narrow}) portion of PLDA training set.
Therefore, we explore extending x-vector training data as well.
Our results indicate, in contrast to back-end, the neural network-based front-end can benefit from synthetic narrowband data.
Further analysis into per-condition results indicate that most benefits come from trials other than CTS-CTS, CTS-AFV, and AFV-AFV.
This observation followed by shifts seen in score histogram reveals a generic calibration-like effect caused by bandwidth extension.
This phenomenon can be further studied in future with more types of trials and stronger x-vector systems.
Our primary choice of test-time scheme is \emph{blind extension}, where we do not detect wideband signals.
Since \emph{SRE21-audio-eval} has some wideband signals, we tested two schemes: skipping extension of wideband test signals (\emph{expand\_narrow}) and not modifying lowerband information (\emph{Lower Frequency Replacement}).
We find that it is beneficial to extend wideband signals as well since (1) it avoids mismatch w.r.t. extended signals and (2) GANs can approximate identity operation.
We also performed a perceptual quality analysis with PESQ, ESTOI, and LSD measures.
We do not observe a positive correlation with downstream performance.
We speculate this is due to the absence of perception-improving loss terms.
However, we visually find GANs to predict upperband information in spectrograms.

One limitation of our work is the degradation in AFV-AFV trial when using the blind extension scheme.
This may be handled via a better identity loss or introducing ASV metrics in BWE training.
We can also investigate deep feature loss~\cite{kataria2021perceptual} and/or self-supervised models to improve perceptual quality as well as downstream performance.
Finally, we encourage the reader to refer to \cite{villalba22b_odyssey,villalba22_odyssey} where we report results on stronger baselines, system fusion, and multi-modal setups.
We extend our work to joint learning with domain adaptation in \cite{kataria2022joint}.
\vspace{-2mm}

\bibliographystyle{IEEEtran}
\bibliography{bare_jrnl.bib}

\clearpage

\appendices
\section{Detailed results for different x-vector models}
\label{sec:appendix}
Here, we detail the results of Table~\ref{tab:detail}.
The results are for three x-vector models.
All models are trained with smaller chunks (4~s) and then fine-tuned on long recordings (10-60~s), per standard verification training procedure.
However, there is a difference in the their training data.
In the first model (Table~\ref{tab:detail1}), x-vector is trained and fine-tuned on original unextended training data (i.e. \emph{wide}, \emph{wide\_down}, and \emph{narrow}).
In the second model (Table~\ref{tab:detail2}), x-vector is trained on unextended data like previous model but is fine-tuned on extended data (i.e. original \emph{wide}, extended \emph{wide\_down}, and extended \emph{narrow}).

\begin{table*}[!b]
\centering
\caption{\label{tab:detail1}
Results when x-vector is trained and fine-tuned (on long recordings) on unextended training data.}
\resizebox{0.96\textwidth}{!}{
\begin{tabular}{@{}lccccc@{}}
\toprule
\textbf{PLDA data} & \textbf{PLDA data extended} & \textbf{Test data extended}  & \textbf{SRE16-YUE-eval40} & \textbf{SRE-CTS-superset-dev} & \textbf{SRE21-audio-eval} \\
\hline
\emph{wide}, \emph{wide\_down}, \emph{narrow} & - & \xmark & 7.12 / 0.376 & 5.36 / 0.216 & 17.12 / 0.644 \\ %
\emph{wide}, \emph{wide\_down}, \emph{narrow} & \emph{wide}, \emph{wide\_down}, \emph{narrow} & \xmark & 7.90 / 0.439 & 6.27 / 0.227 & 17.24 / 0.656 \\ %
\emph{wide}, \emph{wide\_down}, \emph{narrow} & \emph{wide\_down}, \emph{narrow} & \xmark & 7.72 / 0.421 & 5.79 / 0.220 & 16.87 / 0.642 \\ %
\emph{wide}, \emph{wide\_down}, \emph{narrow} & \emph{narrow} & \xmark & 7.90 / 0.409 & 5.50 / 0.217 & 15.28 / 0.613 \\ %
\emph{wide}, \emph{narrow} & \emph{wide}, \emph{narrow} & \xmark & 6.64 / 0.387 & 5.08 / 0.199 & 16.40 / 0.641 \\ %
\emph{wide}, \emph{narrow} & \emph{narrow} & \xmark & 6.55 / 0.372 & 4.54 / 0.189 & 15.22 / 0.621 \\ %
\hline
\emph{wide}, \emph{wide\_down}, \emph{narrow} & - & \cmark & 6.83 / 0.359 & 4.71 / 0.202 & 15.93 / 0.623 \\ %
\emph{wide}, \emph{wide\_down}, \emph{narrow} & \emph{wide}, \emph{wide\_down}, \emph{narrow} & \cmark & 6.57 / 0.370 & 5.02 / 0.207 & 15.71 / 0.617 \\ %
\emph{wide}, \emph{wide\_down}, \emph{narrow} & \emph{wide\_down}, \emph{narrow} & \cmark & 6.45 / 0.357 & 4.91 / 0.205 & 14.98 / 0.605 \\ %
\emph{wide}, \emph{wide\_down}, \emph{narrow} & \emph{narrow} & \cmark & 6.39 / 0.352 & 4.91 / 0.204 & 14.82 / 0.599 \\ %
\emph{wide}, \emph{narrow} & \emph{wide}, \emph{narrow} & \cmark & 5.43 / 0.317 & 4.05 / 0.179 & 15.90 / 0.615 \\ %
\emph{wide}, \emph{narrow} & \emph{narrow} & \cmark & \textbf{5.27} / \textbf{0.307} & \textbf{4.01} / \textbf{0.174} & \textbf{14.33} / \textbf{0.591} \\ %
\bottomrule
\end{tabular}
}
\end{table*}

\begin{table*}[!b]
\centering
\caption{\label{tab:detail2}
Results when x-vector is trained on unextended data but fine-tuned (on long recordings) on extended training data.}
\resizebox{0.96\textwidth}{!}{
\begin{tabular}{@{}lccccc@{}}
\toprule
\textbf{PLDA data} & \textbf{PLDA data extended} & \textbf{Test data extended}  & \textbf{SRE16-YUE-eval40} & \textbf{SRE-CTS-superset-dev} & \textbf{SRE21-audio-eval} \\
\hline
\emph{wide}, \emph{wide\_down}, \emph{narrow} & - & \xmark & 6.88 / 0.366 & 5.42 / 0.219 & 17.56 / 0.650 \\
\emph{wide}, \emph{wide\_down}, \emph{narrow} & \emph{wide}, \emph{wide\_down}, \emph{narrow} & \xmark & 7.57 / 0.402 & 5.41 / 0.218 & 18.72 / 0.671 \\ %
\emph{wide}, \emph{wide\_down}, \emph{narrow} & \emph{wide\_down}, \emph{narrow} & \xmark & 7.39 / 0.391 & 5.33 / 0.218 & 18.30 / 0.657 \\ %
\emph{wide}, \emph{wide\_down}, \emph{narrow} & \emph{narrow} & \xmark & 7.39 / 0.387 & 5.33 / 0.217 & 17.82 / 0.652 \\ %
\emph{wide}, \emph{narrow} & \emph{wide}, \emph{narrow} & \xmark & 6.33 / 0.354 & 4.41 / 0.191 & 17.71 / 0.657 \\ %
\emph{wide}, \emph{narrow} & \emph{narrow} & \xmark & 6.07 / 0.336 & 4.33 / 0.186 & 18.67 / 0.654 \\ %
\hline
\emph{wide}, \emph{wide\_down}, \emph{narrow} & - & \cmark & 6.88 / 0.372 & 5.33 / 0.215 & 15.31 / 0.614 \\ %
\emph{wide}, \emph{wide\_down}, \emph{narrow} & \emph{wide}, \emph{wide\_down}, \emph{narrow} & \cmark & 7.05 / 0.382 & 5.25 / 0.210 & 16.26 / 0.625 \\ %
\emph{wide}, \emph{wide\_down}, \emph{narrow} & \emph{wide\_down}, \emph{narrow} & \cmark & 6.82 / 0.375 & 5.19 / 0.210 & 15.65 / 0.616 \\ %
\emph{wide}, \emph{wide\_down}, \emph{narrow} & \emph{narrow} & \cmark & 6.83 / 0.373 & 5.17 / 0.210 & 15.59 / 0.615 \\ %
\emph{wide}, \emph{narrow} & \emph{wide}, \emph{narrow} & \cmark & 5.50 / 0.327 & 4.19 / 0.181 & 15.66 / 0.611 \\ %
\emph{wide}, \emph{narrow} & \emph{narrow} & \cmark & \textbf{5.43} / \textbf{0.315} & \textbf{4.11} / \textbf{0.175} & \textbf{14.88} / \textbf{0.597} \\ %
\bottomrule
\end{tabular}
}
\end{table*}

\begin{table*}[!b]
\centering
\caption{\label{tab:detail3}
Results when x-vector is trained and fine-tuned (on long recordings) on extended training data.}
\resizebox{0.96\textwidth}{!}{
\begin{tabular}{@{}lccccc@{}}
\toprule
\textbf{PLDA data} & \textbf{PLDA data extended} & \textbf{Test data extended}  & \textbf{SRE16-YUE-eval40} & \textbf{SRE-CTS-superset-dev} & \textbf{SRE21-audio-eval} \\
\hline
\emph{wide}, \emph{wide\_down}, \emph{narrow} & - & \xmark & 7.45 / 0.410 & 5.59 / 0.226 & 18.06 / 0.675 \\
\emph{wide}, \emph{wide\_down}, \emph{narrow} & \emph{wide}, \emph{wide\_down}, \emph{narrow} & \xmark & 8.01 / 0.419 & 5.64 / 0.237 & 20.42 / 0.699 \\ %
\emph{wide}, \emph{wide\_down}, \emph{narrow} & \emph{wide\_down}, \emph{narrow} & \xmark & 7.67 / 0.414 & 5.60 / 0.231 & 19.40 / 0.684 \\ %
\emph{wide}, \emph{wide\_down}, \emph{narrow} & \emph{narrow} & \xmark & 7.55 / 0.410 & 5.92 / 0.238 & 17.25 / 0.657 \\ %
\emph{wide}, \emph{narrow} & \emph{wide}, \emph{narrow} & \xmark & 6.54 / 0.373 & 4.67 / 0.211 & 20.49 / 0.708 \\ %
\emph{wide}, \emph{narrow} & \emph{narrow} & \xmark & 6.25 / 0.371 & 4.42 / 0.200 & 21.81 / 0.697 \\ %
\hline
\emph{wide}, \emph{wide\_down}, \emph{narrow} & - & \cmark & 7.64 / 0.436 & 5.53 / 0.220 & 16.32 / 0.650 \\ %
\emph{wide}, \emph{wide\_down}, \emph{narrow} & \emph{wide}, \emph{wide\_down}, \emph{narrow} & \cmark & 7.40 / 0.425 & 5.35 / 0.220 & 16.49 / 0.643 \\ %
\emph{wide}, \emph{wide\_down}, \emph{narrow} & \emph{wide\_down}, \emph{narrow} & \cmark & 7.26 / 0.419 & 5.31 / 0.218 & 16.28 / 0.642 \\ %
\emph{wide}, \emph{wide\_down}, \emph{narrow} & \emph{narrow} & \cmark & 7.20 / 0.413 & 5.35 / 0.219 & 16.46 / 0.644 \\ %
\emph{wide}, \emph{narrow} & \emph{wide}, \emph{narrow} & \cmark & 5.86 / 0.367 & 4.23 / 0.188 & 15.87 / 0.628 \\ %
\emph{wide}, \emph{narrow} & \emph{narrow} & \cmark & \textbf{5.67} / \textbf{0.363} & \textbf{4.16} / \textbf{0.187} & \textbf{15.29} / \textbf{0.616} \\ %
\bottomrule
\end{tabular}
}
\end{table*}

We use CGAN extension here.
In the third model (Table~\ref{tab:detail3}), x-vector is trained on fine-tuned on extended data (i.e. original \emph{wide}, extended \emph{wide\_down}, and extended \emph{narrow}).
In contrast to Table~\ref{tab:detail}, we here provide results when test set is not extended as well (upper half of three tables).
We make several observations: 1) extending test set is crucial, 2) synthetic narrowband data is harmful for PLDA, and 3) extending wideband data is harmful for PLDA.
We find that the first model brings best performance -- eliminating the possible need for training x-vector on extended data.
In other words, baseline x-vector network need not require re-training.
However, this needs further investigation on larger x-vector architectures like in \cite{villalba22b_odyssey}.

\ifCLASSOPTIONcaptionsoff
  \newpage
\fi

\end{document}